\documentclass[fleqn,usenatbib]{mnras}

\usepackage[utf8]{inputenc}
\usepackage[T1]{fontenc}
\usepackage{newtxtext,newtxmath}
\usepackage{amsmath}
\usepackage{graphicx}
\usepackage{url}
\usepackage{booktabs}
\usepackage{microtype}
\usepackage{longtable}
\usepackage{float}
\usepackage{afterpage}
\usepackage{placeins}
\usepackage{caption}
\usepackage{xcolor} %DIF > 

\newcommand{\sRMS}{\sigma_{\rm rms}}
\newcommand{\DAOPHOT}{\textsc{daophot}}
\newcommand{\ALLSTAR}{\textsc{allstar}}
\newcommand{\Photutils}{\textsc{Photutils}}
\newcommand{\Background}{\textsc{Background2D}}
\newcommand{\MMM}{\textsc{MMMBackground}}

\title[UKIRT M33 VII: 15-yr K-band Database]{%
The UKIRT M33 Monitoring Project \DIFdelbegin \DIFdelend \DIFaddbegin \DIFadd{-- }\DIFaddend VII. A 15-year K-band 
photometric catalogue of long-period variable stars 
in the central kiloparsec of M33}

\author[M. Alizadeh et al.]{
Mina Alizadeh,$^{1}$
Atefeh Javadi,$^{2}$\thanks{Corresponding author. E-mail: atefeh@ipm.ir}
Yousefali Abedini,$^{1,3}$
\newauthor
Fatemeh Nikzat,$^{2}$
Hedieh Abdollahi,$^{2,4}$
and Jacco Th. van Loon$^{5}$
\\
$^{1}$Department of Physics, Faculty of Science,
University of Zanjan, 38791-45371, Zanjan, Iran\\
$^{2}$School of Astronomy, Institute for Research in Fundamental Sciences
(IPM), Tehran, 19568-36613, Iran\\
$^{3}$Center for Research in Climate Change and Global Warming
(CRCC), IASBS, Zanjan, Iran\\
$^{4}$Konkoly Observatory, HUN-REN Research Centre for Astronomy and
Earth Sciences, Konkoly-Thege Mikl\'os \'ut 15--17,
H-1121 Budapest, Hungary\\
$^{5}$Lennard-Jones Laboratories, Keele University,
Keele, Staffordshire ST5 5BG, UK
}
\pubyear{2026}
\DeclareOldFontCommand{\bf}{\normalfont\bfseries}{\mathbf} %DIF PREAMBLE
\providecommand{\DIFadd}[1]{#1} %DIF PREAMBLE
\providecommand{\DIFdel}[1]{} %DIF PREAMBLE
\providecommand{\DIFaddbegin}{} %DIF PREAMBLE
\providecommand{\DIFaddend}{} %DIF PREAMBLE
\providecommand{\DIFdelbegin}{} %DIF PREAMBLE
\providecommand{\DIFdelend}{} %DIF PREAMBLE
\providecommand{\DIFaddFL}[1]{\DIFadd{#1}} %DIF PREAMBLE
\providecommand{\DIFdelFL}[1]{} %DIF PREAMBLE
\providecommand{\DIFaddbeginFL}{} %DIF PREAMBLE
\providecommand{\DIFaddendFL}{} %DIF PREAMBLE
\providecommand{\DIFdelbeginFL}{} %DIF PREAMBLE
\providecommand{\DIFdelendFL}{} %DIF PREAMBLE
\RequirePackage{listings} %DIF PREAMBLE
\lstdefinelanguage{DIFcode}{ %DIF PREAMBLE
  moredelim=[il][\color{white}\tiny]{\%DIF\ <\ }, %DIF PREAMBLE
  moredelim=[il][\sffamily]{\%DIF\ >\ } %DIF PREAMBLE
} %DIF PREAMBLE
\lstdefinestyle{DIFverbatimstyle}{ %DIF PREAMBLE
	language=DIFcode, %DIF PREAMBLE
	basicstyle=\ttfamily, %DIF PREAMBLE
	columns=fullflexible, %DIF PREAMBLE
	keepspaces=true %DIF PREAMBLE
} %DIF PREAMBLE
\lstnewenvironment{DIFverbatim}{\lstset{style=DIFverbatimstyle}}{} %DIF PREAMBLE
\lstnewenvironment{DIFverbatim*}{\lstset{style=DIFverbatimstyle,showspaces=true}}{} %DIF PREAMBLE

\begin{document}
\label{firstpage}
\pagerange{\pageref{firstpage}--\pageref{lastpage}}
\maketitle

%%
%% ============================================================
%% ABSTRACT
%% ============================================================
%%
\begin{abstract}
We present a homogenised K-band photometric catalogue for the
central \DIFdelbegin \DIFdelend \DIFaddbegin \DIFadd{${\approx}12$}\DIFaddend ~arcmin$^2$ (\DIFdelbegin \DIFdelend \DIFaddbegin \DIFadd{${\approx}0.9$}\DIFaddend ~kpc$^2$\DIFdelbegin \DIFdelend ) of the
Triangulum spiral galaxy M33, combining four
independent PSF-based near-infrared \DIFdelbegin \DIFdelend \DIFaddbegin \DIFadd{datasets }\DIFaddend obtained with UKIRT
between 2003 and \DIFdelbegin \DIFdelend \DIFaddbegin \DIFadd{2018. These comprise }\DIFaddend the 2003--2007 UIST
multi-epoch monitoring survey\DIFdelbegin \DIFdelend \DIFaddbegin \DIFadd{, }\DIFaddend the 2005 UFTI K-band mosaic \DIFdelbegin \DIFdelend \DIFaddbegin \DIFadd{(}\DIFaddend for
which we present new \DIFdelbegin \DIFdelend PSF \DIFdelbegin \DIFdelend \DIFaddbegin \DIFadd{photometry), new WFCAM PSF
reductions }\DIFaddend for the four central tiles\DIFdelbegin \DIFdelend \DIFaddbegin \DIFadd{, }\DIFaddend and a single K-band epoch
from the 2018 UKIRT Hemisphere Survey (UHS~DR3). Cross-matching all
four catalogues \DIFdelbegin \DIFdelend and retaining stars detected in at least two of
\DIFdelbegin \DIFdelend \DIFaddbegin \DIFadd{them yields }\DIFaddend 847~unique stars comprising 21\,085~valid K-band
measurements \DIFdelbegin \DIFdelend spanning a baseline of 15.11~yr\DIFdelbegin \DIFdelend . Photometric
homogenisation onto the 2MASS/Vega system via 82~\DIFdelbegin \DIFdelend \DIFaddbegin \DIFadd{low-scatter
}\DIFaddend calibration stars removes systematic inter-dataset offsets of up to
0.22~mag. Of the 847~monitored stars, 771 (91~per cent) are
classified as variable ($\sigma_{\rm rms} > 0.10$~mag), including
385~high-amplitude variables ($A_K \geq 1.0$~mag)\DIFdelbegin \DIFdelend \DIFaddbegin \DIFadd{: }\DIFaddend prime Mira-type
long-period variable candidates \DIFdelbegin \DIFdelend \DIFaddbegin \DIFadd{for period determination}\DIFaddend .
Cross-matching with archival \textit{Spitzer} IRAC photometry
yields 156~matched stars, of which 120 satisfy $\langle K\rangle -
[3.6] > 0.5$~mag, identifying dust-enshrouded AGB candidates
inaccessible to \DIFdelbegin \DIFdelend all-sky optical surveys. This \DIFdelbegin \DIFdelend catalogue constitutes
the most extensive near-infrared photometric baseline assembled for
AGB stars in any spiral galaxy beyond the Milky Way, \DIFdelbegin \DIFdelend \DIFaddbegin \DIFadd{providing }\DIFaddend the
photometric foundation for \DIFaddbegin \DIFadd{the }\DIFaddend period-luminosity analysis and
\DIFdelbegin \DIFdelend \DIFaddbegin \DIFadd{star-formation-history reconstruction }\DIFaddend in the companion Paper~VIII.
The full catalogue is available at the Centre de Donn\'{e}es
astronomiques de Strasbourg (CDS).
\end{abstract}

\begin{keywords}
stars: AGB and post-AGB \DIFdelbegin \DIFdelend \DIFaddbegin \DIFadd{--
}\DIFaddend stars: long-period variables \DIFdelbegin \DIFdelend \DIFaddbegin \DIFadd{--
}\DIFaddend stars: variables: other \DIFdelbegin \DIFdelend \DIFaddbegin \DIFadd{--
}\DIFaddend galaxies: individual: M33 \DIFdelbegin \DIFdelend \DIFaddbegin \DIFadd{--
}\DIFaddend galaxies: stellar content \DIFdelbegin \DIFdelend \DIFaddbegin \DIFadd{--
}\DIFaddend catalogues
\end{keywords}
%%
%%%% ============================================================
%% SECTION 1 --- INTRODUCTION
%% ============================================================
%%
\section{Introduction}
\label{sec:intro}

Reliable period determination for long-period variables (LPVs) on the
asymptotic giant branch (AGB) demands photometric time series spanning
years to decades.
Mira variables and semi-regular pulsators have characteristic periods of
$P \sim 100$--2000~d \citep{Wood2000, Ita2004a, 2019ApJ...877...49G, 2026ApJ..1000...69A}, so a temporal baseline
of only 3--5~yr samples at most two or three pulsation cycles for the
longest-period objects, insufficient to resolve period aliases, detect
secular period changes, or classify borderline semi-regular behaviour
\citep{Soszynski2004, Soszynski2009} with confidence.
Yet it is precisely the period distribution of the AGB population, through
the period-luminosity (P-L) relation
\citep{Feast1989, Ita2004a, Whitelock2008, Soszynski2009},
that underpins distance determination, luminosity function construction,
and ultimately the reconstruction of a galaxy's star formation history
\citep[SFH;][]{Javadi2013, Javadi2017}.
The scientific return from any near-infrared variability survey is therefore
fundamentally limited by the length and homogeneity of its photometric
baseline.

Constructing such a baseline across a decade-long archive is not
straightforward. Independent observing campaigns use different
instruments, detector formats, PSF-fitting codes, and photometric
zero-point conventions, producing systematic inter-dataset offsets
that, if uncorrected, masquerade as intrinsic variability and bias
period estimates. The problem is compounded when aperture-based and
PSF-based reductions are mixed\DIFdelbegin \DIFdelend \DIFaddbegin \DIFadd{, since aperture photometry in crowded
fields }\DIFaddend suffers from magnitude-dependent blending biases
\DIFdelbegin \DIFdelend \DIFaddbegin \DIFadd{(Section~\ref{sec:aperture_limitations})}\DIFaddend . Building a homogenised,
calibration-verified photometric catalogue from heterogeneous
archival data is therefore a necessary prerequisite for robust
period analysis.

At near-infrared (NIR) wavelengths, and in the K band (2.2~$\mu$m) in
particular, these challenges are offset by decisive observational
advantages.
AGB stars and red supergiants (RSGs) are detected with minimum
circumstellar extinction and maximum contrast against the stellar
background; their K-band pulsation amplitudes ($A_K \sim 0.2$--2~mag)
far exceed typical ground-based photometric uncertainties
\citep{Wood1992, Wood1998, Whitelock2003}.
The K band falls in the Rayleigh--Jeans tail of cool photospheres
($T_{\rm eff} \approx 2500$--4000~K) but shortward of the peak of warm
circumstellar dust emission ($T_{\rm dust} \approx 500$--1000~K;
\citealt{Feast1989, Javadi2011a}), making it the optimal window for
identifying and monitoring bright, dusty, large-amplitude variables.
Both thermally pulsing AGB stars and RSGs manifest as LPVs with periods
of 150--2000~d \citep{Wood2000, Ita2004a} and are expected to dominate
the bright NIR population of any star-forming galaxy with a composite
stellar population.
AGB and red supergiant mass-loss of $10^{-8}$--$10^{-4}$~$M_\odot$~yr$^{-1}$
\citep{Vassiliadis1993, vanLoon1999, Hofner2018, vanLoon2025, 2026ApJ..1006..131G} return up to 85~per cent
of their initial mass, enriched in carbon, nitrogen, and $s$-process
elements \citep{Herwig2005, Marigo2008, 2017ApJ...851..152B, 2025ApJ...991...24B}, to the interstellar medium, making them indispensable tracers of stellar
evolution and galactic chemical enrichment.

Among nearby galaxies, the Triangulum spiral M33 provides an
irreplaceable laboratory for these studies.
At \DIFdelbegin \DIFdelend \DIFaddbegin \DIFadd{${\approx}955$}\DIFaddend ~kpc \citep[$\mu = 24.9$~mag;][]{Bonanos2006, Javadi2011a}, independently
confirmed by \citet{Taormina2026}, and with a moderate
inclination of $i \approx 56\degr$ \citep{Deul1987}, individual evolved
stars can be resolved and photometrically monitored across the full
optical disc in the NIR. M31 offers a comparable opportunity, though at a
higher inclination that complicates such resolved photometry; more
importantly, JWST now extends this capability to galaxies at distances
of a few~Mpc.
Unlike the genuinely metal-poor environments of the Small and Large
Magellanic Clouds ($Z \approx 0.2$--$0.5\,Z_\odot$;
\citealt{Russell1992}), M33 exhibits near-solar metallicity in its inner
disc with a radial gradient of ${\approx}-0.05$~dex~kpc$^{-1}$
\citep{Rosolowsky2008, Bresolin2011}.
This metallicity structure directly modulates dust condensation
efficiency, grain species, and wind-driving mechanisms in evolved stars
\citep{Ventura2012, Nanni2019}, making M33 the only galaxy in which
metallicity-dependent AGB properties can be studied statistically against
a large ($N \gtrsim 10^3$), spatially resolved, multi-epoch NIR sample at
near-solar metallicity \citep{Williams2024}.
The method of reconstructing star formation histories from LPV luminosity
functions, pioneered \DIFdelbegin \DIFdelend \DIFaddbegin \DIFadd{in Paper~I \citep{Javadi2011a} }\DIFaddend and subsequently applied to
M33 \citep{Javadi2011b, Javadi2013} and to numerous Local Group galaxies
\citep{Rezaeikh2014, Hashemi2019, Saremi2020, Navabi2021, Abdollahi2023, 2026arXiv260934329K}, relies critically on having a
photometrically homogeneous sample of AGB variables.

The United Kingdom Infrared Telescope (UKIRT) has accumulated an
exceptional time-series archive for M33 since 2003, combining dedicated
multi-epoch monitoring with the UKIRT Imager Spectrometer
\citep[UIST;][]{RamsayHowat2004}, deep single-epoch imaging with the
UKIRT Fast Track Imager \citep[UFTI;][]{Roche2003}, wide-field coverage
with the Wide Field Camera \citep[WFCAM;][]{Casali2007}, and most
recently a single deep epoch from the UKIRT Hemisphere Survey
\citep[UHS~DR3;][]{Schneider2025}.
Together these instruments span 15.1~yr (2003--2018) in the K band,
constituting one of the longest and deepest NIR variability baselines
available for any metal-rich galaxy beyond the Milky Way.

Earlier papers in this series have progressively exploited subsets of
this archive.
\DIFdelbegin \DIFdelend Paper~I \DIFdelbegin \DIFdelend \DIFaddbegin \DIFadd{\citep{Javadi2011a} }\DIFaddend established the monitoring
programme and presented the first K-band variability catalogue for
the central \DIFdelbegin \DIFdelend \DIFaddbegin \DIFadd{$4'\times4'$ field }\DIFaddend (\DIFdelbegin \DIFdelend 2003--2007 UIST campaign\DIFdelbegin \DIFdelend \DIFaddbegin \DIFadd{, }\DIFaddend baseline
3.80~yr)\DIFdelbegin \DIFdelend \DIFaddbegin \DIFadd{. }\DIFaddend Paper~II \DIFdelbegin \DIFdelend \DIFaddbegin \DIFadd{\citep{Javadi2011b} }\DIFaddend derived the star formation
history of the central kiloparsec \DIFdelbegin \DIFdelend \DIFaddbegin \DIFadd{from the resulting LPV census.
}\DIFaddend Paper~III \DIFdelbegin \DIFdelend \DIFaddbegin \DIFadd{\citep{Javadi2013} }\DIFaddend derived star formation and mass-return
rates\DIFdelbegin \DIFdelend , providing the first quantitative estimate of AGB dust input
to the M33 ISM\DIFdelbegin \DIFdelend \DIFaddbegin \DIFadd{. }\DIFaddend Paper~IV \DIFdelbegin \DIFdelend \DIFaddbegin \DIFadd{\citep{Javadi2015} extended the census to
the full disc via }\DIFaddend wide-field \DIFdelbegin \DIFdelend \DIFaddbegin \DIFadd{WFCAM aperture photometry
(}\DIFaddend 4643~variable red \DIFdelbegin \DIFdelend \DIFaddbegin \DIFadd{giants); this aperture catalogue is excluded from
the present work for reasons discussed in
Section~\ref{sec:aperture_limitations}. }\DIFaddend Paper~V \DIFdelbegin \DIFdelend \DIFaddbegin \DIFadd{\citep{Javadi2017}
}\DIFaddend constructed the stellar birth mass function and \DIFdelbegin \DIFdelend the spatially
resolved SFH across the full \DIFdelbegin \DIFdelend \DIFaddbegin \DIFadd{disc. }\DIFaddend Paper~VI
\DIFdelbegin \DIFdelend \DIFaddbegin \DIFadd{\citep{javadi2026ukinfraredtelescopem33} }\DIFaddend measured dust and gas
mass-loss rates for AGB stars and RSGs\DIFdelbegin \DIFdelend , combining near-IR
variability with mid-IR SED modelling\DIFdelbegin \DIFdelend .
Throughout, however, three compounding limitations have prevented full
exploitation of the combined archive.
First, the four independent observing campaigns (UIST, UFTI, WFCAM, UHS)
were reduced with heterogeneous pipelines and zero-point conventions,
producing systematic inter-dataset offsets of up to 0.22~mag that, if
uncorrected, bias period estimates and variability statistics.
Second, the WFCAM photometry of Paper~IV is aperture-based \DIFdelbegin \DIFdelend \DIFaddbegin \DIFadd{and
suffers from blending biases }\DIFaddend in the crowded nuclear field
(\DIFdelbegin \DIFdelend \DIFaddbegin \DIFadd{Section~\ref{sec:aperture_limitations}), }\DIFaddend rendering those
measurements unsuitable for precise light-curve construction.
Third, with only the 3.80-yr UIST baseline, period determination for
Mira variables with $P \gtrsim 500$~d was limited to at most 2--3
pulsation cycles \DIFdelbegin \DIFdelend \DIFaddbegin \DIFadd{(see above)}\DIFaddend .

Here we address all three limitations simultaneously \DIFdelbegin \DIFdelend \DIFaddbegin \DIFadd{by }\DIFaddend extending
the baseline from 3.80 \DIFdelbegin \DIFdelend to 15.11~yr \DIFdelbegin \DIFdelend \DIFaddbegin \DIFadd{through three new contributions}\DIFaddend :
(i)~\DIFdelbegin \DIFdelend PSF photometry for the 2005 UFTI K-band mosaic\DIFdelbegin \DIFdelend ,
presented here for the first time; (ii)~new PSF reductions of WFCAM
imaging for the four central tiles \DIFdelbegin \DIFdelend \DIFaddbegin \DIFadd{\citep{Alizadeh2025}}\DIFaddend , superseding
the aperture photometry of Paper~IV; and (iii)~K-band PSF photometry of the 2018
UHS epoch\DIFdelbegin \DIFdelend , providing the long-baseline anchor essential for period
determination of the longest-period AGB variables. \DIFdelbegin \DIFdelend \DIFaddbegin \DIFadd{The resulting
homogenised catalogue and its properties are summarised in
Section~\ref{sec:results}}\DIFaddend .
Period determinations, period-luminosity relations, and SFH
reconstruction for this sample are presented in a companion paper
(Paper~VIII).

The paper is organised as follows.
Section~\ref{sec:obs} describes the five datasets and their PSF
photometry pipelines, including validation of the \textsc{Photutils}
pipeline against archival \textsc{daophot} photometry.
Section~\ref{sec:photcat} presents the cross-matching procedure and
photometric homogenisation.
Section~\ref{sec:variables} describes the variability detection and
amplitude classification.
Section~\ref{sec:results} presents the catalogue properties, including
the K-band $\sigma_K$ distribution, amplitude--magnitude relation, and
multi-wavelength colour analysis.
Section~\ref{sec:conclusions} \DIFdelbegin \DIFdelend \DIFaddbegin \DIFadd{summarises }\DIFaddend the principal results.
The full homogenised catalogue will be made available at the
Centre de Donn\'ees astronomiques de Strasbourg (CDS) upon acceptance.
\section{Observations and data}
\label{sec:obs}

Our analysis draws on five independent near-infrared datasets obtained with
UKIRT between 2003 and 2018.
Four are used as PSF-based catalogues in the combined light-curve
construction: the 2003--2007 UIST multi-epoch monitoring, the 2005 UFTI
K-band mosaic, PSF reductions of WFCAM data for four central tiles, and
the 2018 UHS K-band epoch.
The fifth dataset, the wide-field aperture photometry published in
Paper~IV \citep{Javadi2015}, is described here for completeness. Its
aperture-based catalogue is excluded from the homogenised catalogue owing
to blending biases in the crowded nuclear field (Section~\ref{sec:wfcam});
the same WFCAM observations are instead retained via the new PSF-based
reduction presented in this work.
Together the four PSF catalogues extend the temporal baseline from
3.80~yr (UIST alone) to 15.11~yr, enabling period determinations for
variables with $P \lesssim 2000$~d.
Key properties are summarised in Table~\ref{tab:datasets}.

\begin{table*}
\centering
\caption{Summary of near-infrared datasets used in this work.}
\label{tab:datasets}
\begin{tabular}{llccccrl}
\hline
Instrument & Dates & HJD range & $N_{\rm ep}$ & Filters
           & FoV (arcmin$^{2}$) & $N_{\rm src}$ & Status \\
           & & ($-$2\,450\,000) & & & & & \\
\hline
UIST       & 2003--2007 & 2919--4307 & 23\,$K$ / 5\,$J$ / 2\,$H$
           & $J,H,K$ & \DIFdelbeginFL \DIFdelendFL \DIFaddbeginFL \DIFaddFL{${\sim}16$ }\DIFaddendFL & 18\,398 & Published (Paper~I) \\
UFTI       & 2005 Aug   & 3594       & 1$^{a}$
           & $K$     & ${\sim}16$ & 18\,509 & This work (PSF) \\
WFCAM-PSF  & 2005--2007 & 3631--4397 & 5--8\,($K$)$^{b}$
           & $J,H,K$ & ${\sim}1000$ & 211\,179 & This work; \citealt{Alizadeh2025} \\
WFCAM-Aper & 2005--2007 & 3631--4398 & multiple
           & $J,H,K$ & ${\sim}3\,200$ & 403\,734 & Published (Paper~IV); not used$^{c}$ \\
UHS DR3    & 2018 Nov   & 8438       & 1
           & $K$     & Survey & 2\,826$^{d}$ & This work \\
\hline
\end{tabular}
\begin{minipage}{\linewidth}
\smallskip
\textit{Notes.}
($a$)~UFTI spans three consecutive nights (2005 Aug 11--13); treated as
one epoch.
($b$)~K-band epochs by tile: Q1: 8; Q2: 8 (longest baseline, 731~d);
Q3: 6; Q4: 5. See Table~\ref{tab:wfcam_log}.
($c$)~Excluded from the homogenised light-curve catalogue owing to blending
biases exceeding 0.5~mag for ${\sim}3.5$~per cent of sources within 5~arcmin
of the nucleus; see Section~\ref{sec:wfcam} for details.
($d$)~Sources within a 3-arcmin radius of the M33 nucleus with flags $= 0$.
Abbreviations: $N_{\rm ep}$, number of epochs; FoV, total field of view;
$N_{\rm src}$, catalogued sources.
References: Paper~I $=$ \citet{Javadi2011a}; Paper~IV $=$ \citet{Javadi2015};
UHS~DR3 $=$ \citet{Schneider2025}.
\end{minipage}
\end{table*}

%% ---- 2.1 UIST ----
\subsection{UIST multi-epoch monitoring (2003--2007)}
\label{sec:uist}

The temporal foundation of our variability study is the K-band monitoring
programme conducted with UIST \citep{RamsayHowat2004} between 2003 October
and 2007 July.
The $1024\times1024$~pixel InSb array provides a plate scale of
0.12~arcsec~pixel$^{-1}$\DIFdelbegin \DIFdelend \DIFaddbegin \DIFadd{; four overlapping quadrants cover a
${\approx}4\times4$~arcmin }\DIFaddend field of view
centred on the M33 nucleus
($\alpha = 01^{\rm h}\,33^{\rm m}\,50\rlap{.}^{\rm s}9$,
$\delta = +30\degr\,39\arcmin\,37\arcsec$, J2000).
Over 23 distinct K-band observing nights (50 individual exposures,
HJD~2\,452\,919--2\,454\,307; baseline 1388~d $= 3.80$~yr) the programme
accumulated a median of 12 epochs per star.
Supplementary $J$ (5 nights) and $H$ (2 nights) observations were obtained
on a subset of nights.

PSF photometry with \DAOPHOT/\ALLSTAR\ \citep{Stetson1987} yielded a
catalogue of 18\,398 sources calibrated against 2MASS \citep{Skrutskie2006}.
Typical K-band precision: $\sigma_K = 0.023$~mag (median, $K < 17$~mag),
rising to ${\sim}0.037$~mag at $17 < K < 18$~mag.
Variability was characterised with the \citet{Stetson1996} $L$-index;
5\,648 sources satisfy $L > 1.5$ and 12\,367 satisfy $L > 0.7$.
\DIFdelbegin \DIFdelend We adopt the UIST catalogue as the primary astrometric reference frame
for all cross-matching.

%% ---- 2.2 UFTI ----
\subsection{UFTI deep K-band mosaic (2005 August)}
\label{sec:ufti}

Deep K-band imaging was obtained with UFTI \citep{Roche2003} over three
consecutive nights (2005 August 11--13; HJD~2\,453\,594).
The $1024\times1024$~pixel HgCdTe array (plate scale
0.091~arcsec~pixel$^{-1}$; field $92\rlap{.}''9 \times 92\rlap{.}''9$)
was pointed in a $3\times3$ mosaic with slight overlap, covering
${\approx}16$~arcmin$^2$ centred on the M33 nucleus.
Thirty K-band frames were acquired and co-added after standard
near-infrared reduction (non-linearity correction, flat-fielding, sky
subtraction, 2MASS-based astrometry).
PSF photometry via \DAOPHOT/\ALLSTAR\ yielded 18\,509 detections to
$K \approx 18.5$~mag (median $\sigma_K \approx 0.04$~mag for $K < 18$~mag).
Calibration against UIST on a 1-arcsec radius yields ${\approx}13\,000$
common sources.
The three-night span of 2.4~d is negligible relative to LPV periods;
the mosaic is treated as a single photometric epoch.
The image was published in Paper~I; the PSF reductions \DIFdelbegin \DIFdelend \DIFaddbegin \DIFadd{follow
}\DIFaddend \citet{Nikzat2012} and are published in final homogenised form here
for the first time.

%% ---- 2.3 WFCAM ----
\subsection{WFCAM wide-field monitoring (2005--2007)}
\label{sec:wfcam}

\subsubsection{Instrument and coverage}

WFCAM \citep{Casali2007} observed M33 between 2005 September and 2007
October in four tiles (Q1--Q4, following Paper~IV), centred at:
Q1 ($01^{\rm h}33^{\rm m}19^{\rm s}$, $+30\degr32'50''$);
Q2 ($01^{\rm h}34^{\rm m}22^{\rm s}$, $+30\degr32'50''$);
Q3 ($01^{\rm h}34^{\rm m}22^{\rm s}$, $+30\degr46'23''$);
Q4 ($01^{\rm h}33^{\rm m}19^{\rm s}$, $+30\degr46'23''$; J2000).
Each $2048\times2048$~pixel Rockwell Hawaii-II detector subtends
${\approx}13\rlap{.}'65\times13\rlap{.}'65$ at 0.4~arcsec~pixel$^{-1}$;
the four tiles together cover ${\sim}0.89$~deg$^2$, overlapping the UIST
and UFTI fields in the central ${\sim}16$~arcmin$^2$.
The complete observing log is in Table~\ref{tab:wfcam_log}.

\subsubsection{Aperture photometry (Paper~IV) and its limitations}
\DIFdelbegin %

\DIFdelend \DIFaddbegin \label{sec:aperture_limitations}
\DIFaddend Paper~IV published aperture photometry of the full WFCAM
survey (403\,734~sources), processed through the CASU pipeline.
In the crowded central region ($r < 5$~arcmin), discrepancies
$\Delta m > 0.5$~mag affect ${\sim}3.5$~per cent of sources, with
78~per cent of outliers within 5~arcmin of the nucleus.
This aperture catalogue is therefore excluded from our combined light-curve
analysis and is not included in the homogenised Paper~VII catalogue.
The PSF-based WFCAM catalogue described below supersedes it for science
requiring photometric accuracy in the nuclear field. 

\begin{figure*}
  \includegraphics[width=\textwidth]{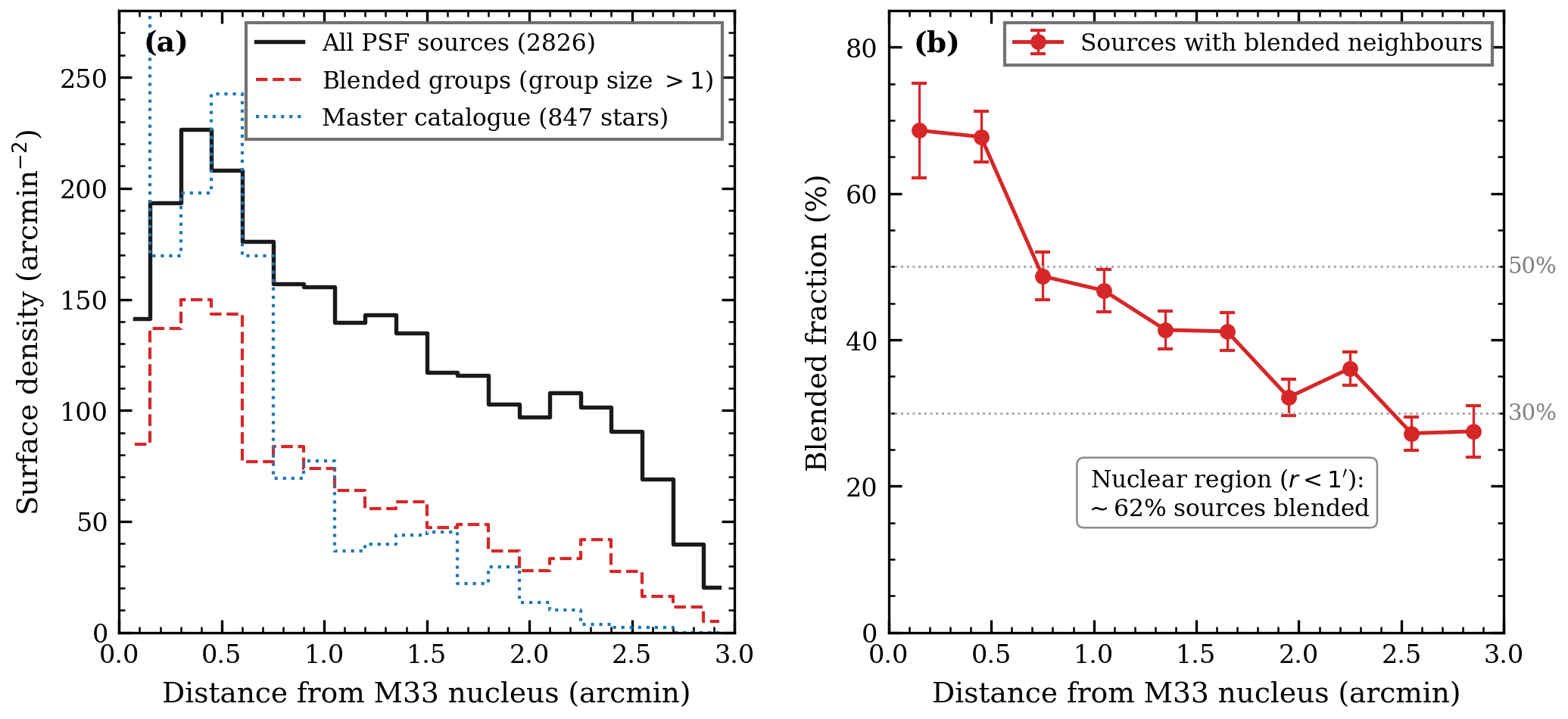}
  \caption{Stellar crowding in the M33 nuclear field from UHS
  $K$-band PSF photometry.
  \textit{(a)}~Radial surface density profiles for all 2826~PSF
  sources (solid black), sources in blended PSF groups of two or
  more stars (dashed red), and the 847-star master catalogue of
  this work (dotted blue).
  \textit{(b)}~Fraction of sources belonging to blended PSF groups
  as a function of distance from the M33 nucleus, with binomial
  uncertainties.
  The blended fraction reaches ${\sim}62$~per cent at
  $r < 1$~arcmin, demonstrating that aperture photometry is
  unreliable for precise light-curve construction in the nuclear
  field; this motivates the exclusive use of PSF photometry in
  this work \citep[see also][]{Alizadeh2024}.}
  \label{fig:crowding}
\end{figure*}

The magnitude-dependent blending biases between aperture and PSF photometry in the M33 nuclear field have been quantified directly in \citet{Alizadeh2024}\DIFdelbegin \DIFdelend .

\subsubsection{New PSF photometry}
\label{sec:wfcam_psf}

PSF photometry was independently performed on all WFCAM frames for the
four central tiles using \DAOPHOT/\ALLSTAR\ \citep{Stetson1987}.
The reductions employed an empirical PSF model constructed from isolated
bright stars in each individual frame, following identical procedures to
the UIST and UFTI reductions \citep{Nikzat2012}.
Cross-matching of the resulting single-epoch catalogues across all four tiles
and all epochs \DIFdelbegin \DIFdelend \DIFaddbegin \DIFadd{yielded }\DIFaddend a unified PSF catalogue of 211\,179~unique
sources across four tiles (see also \citealt{Alizadeh2025}).
The K-band temporal baseline varies by tile: Q2 has 8~epochs spanning
731~d; Q1 has 7~epochs; Q3 has 6~epochs; Q4 has 5~epochs
(Table~\ref{tab:wfcam_log}).

%% ---- 2.4 UHS ----
\subsection{UHS K-band epoch (2018 November)}
\label{sec:uhs}

\subsubsection{Data acquisition}

A single deep K-band epoch was obtained from UHS~DR3
\citep{Schneider2025} via the WFCAM Science Archive
\citep[WSA;][]{Hambly2008}.
We selected Multiframe~10678650 (Detector~3) from 2018 November~16,
in which the M33 nucleus lies only 0.1~arcmin from the detector centre;
the full UFTI science field (3-arcmin radius) lies well within the
detector with margins exceeding 3.4~arcmin.
Observing conditions are listed in Table~\ref{tab:uhs_obs}.

\begin{table}
\centering
\caption{UHS K-band epoch properties
         (Multiframe~10678650, Detector~3).}
\label{tab:uhs_obs}
\begin{tabular}{ll}
\hline
Parameter & Value \\
\hline
Survey / release                   & UHS DR3 \citep{Schneider2025} \\
Multiframe ID                      & 10678650 (Detector~3) \\
Observation date                   & 2018 November 16 \\
HJD ($-$2\,450\,000)               & 8438.50 \\
Airmass                            & 1.039 \\
Seeing (FWHM)                      & 1.01~arcsec (2.52~pixels) \\
PSF ellipticity                    & 0.94 \\
Sky background                     & $7082 \pm 173$~counts (S/N~$= 40.9$) \\
Pixel scale                        & 0.402~arcsec~pixel$^{-1}$ \\
Detection threshold                & $5\sigma$ \\
PSF model                          & IntegratedGaussianPRF
                                     ($\sigma = 1.06$~pix) \\
Sources (full detector)            & 10\,546 \\
PSF success rate (full detector)   & 96.7 per cent \\
Sources in UFTI field ($r < 3'$)   & 2\,826 \\
PSF success rate (UFTI field)      & 97.1 per cent \\
$K$ range (UFTI field)             & 11.92--18.47~mag \\
Median $K$ (UFTI field)            & 17.09~mag \\
Median $\sigma_K$ ($K < 15$)       & 0.018~mag \\
Median $\sigma_K$ ($15 \le K < 16$)& 0.029~mag \\
Median $\sigma_K$ ($16 \le K < 17$)& 0.061~mag \\
Median $\sigma_K$ ($17 \le K < 18$)& 0.099~mag \\
Astrometric accuracy               & ${\sim}0.1$~arcsec rms \\
\hline
\end{tabular}
\end{table}

\subsubsection{PSF photometry}

PSF photometry was performed with \Photutils\ v1.13 \citep{Bradley2023},
a Python library within the Astropy ecosystem
\citep{AstropyCollaboration2022}.
\Photutils\ implements the \citet{Stetson1987} source-detection and
iterative PSF-fitting algorithms; \textsc{DAOStarFinder} applies the DAOFIND
kernel \citep{Stetson1987}, and \textsc{IterativePSFPhotometry} performs
iterative subtraction of fitted sources following an identical mathematical
framework to \DAOPHOT/\ALLSTAR\ \citep{Bradley2023}.
The software is actively maintained by the Astropy community, is routinely
used in the official reduction pipelines for \textit{Hubble Space Telescope}
and \textit{James Webb Space Telescope} data \citep{AstropyCollaboration2022},
and is cited in peer-reviewed publications for PSF photometry of crowded
extragalactic fields \citep{AstropyCollaboration2022}.

Sky background was modelled with \Background\ using the \MMM\ estimator
\DIFdelbegin \DIFdelend \DIFaddbegin \DIFadd{\citep{Stetson1987} }\DIFaddend on a $50\times50$-pixel mesh (bicubic interpolation;
post-subtraction residual rms $= 21.19$~counts).
Sources were detected with \textsc{DAOStarFinder} at $5\sigma$ with
morphological filters (sharpness $\in [0.2, 1.0]$;
$|\text{roundness}| \le 1.0$), yielding 10\,546~detections.
Stellar profiles were modelled with an IntegratedGaussianPRF
($\sigma = 1.06$~pix; FWHM~$= 1.01$~arcsec).
Sources within $2.5\times\text{FWHM} = 6.25$~pixels of a neighbour were
grouped and fitted simultaneously via Levenberg--Marquardt optimisation in
$11\times11$-pixel windows, iterated three times with re-detection of faint
companions in PSF-subtracted residuals \citep{Stetson1987}.

Within the 3-arcmin UFTI field, PSF photometry yields 2\,826~sources with
flags~$= 0$ (surface density 99.9~arcmin$^{-2}$), 17~per cent more than the
${\sim}2\,400$ recovered by Paper~I, attributable to
the superior 2018 seeing (FWHM~$= 1.01$~arcsec vs.\ 1.2--1.5~arcsec in
the earlier campaign) and iterative deblending.

\subsubsection{Validation of the \textsc{Photutils} pipeline}
\label{sec:validation}

To verify the fidelity of the \Photutils\ PSF-photometry pipeline prior to
its application to the UHS K-band science data, we performed a direct,
pixel-level comparison against the archival \DAOPHOT/\ALLSTAR\ photometry of
Paper~I for the UFTI H-band stacked image from epoch 2005
January 18.
The use of an H-band frame reflects a practical constraint: the H-band
FITS image was accessible for independent reprocessing, whereas the
original K-band FITS frames from Paper~I are no longer available in
the group's data archive.
Nevertheless, the test is rigorous, because the PSF-fitting procedure is
wavelength-independent by construction.

We ran the \Photutils\ pipeline on this image with H-band PSF parameters
(FWHM~$= 6.7$~pixel; \texttt{fit\_shape}~$= 11\times11$~pixel), yielding
7\,448~sources.
Cross-matching against the \DAOPHOT\ catalogue within 1~arcsec returned
1\,522~common sources with a median astrometric separation of
0.14~arcsec, confirming reliable positional correspondence.

\begin{figure*}
  \centering
  \includegraphics[width=\textwidth]{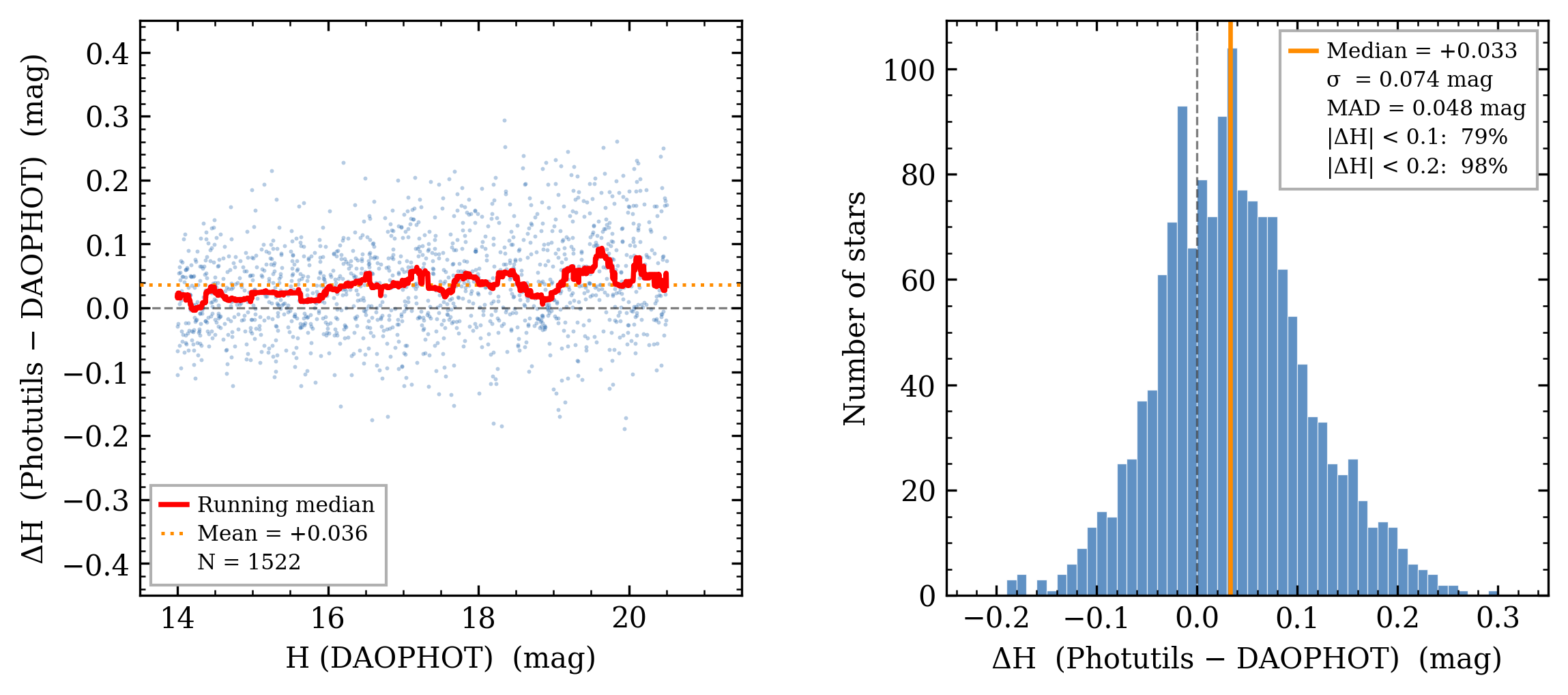}
  \caption{Validation of the \textsc{Photutils} PSF-photometry pipeline
  against \textsc{daophot}/\textsc{allstar} on the UFTI H-band epoch
  2005 January~18 ($N = 1522$ matched sources within 1~arcsec;
  \citealt{Javadi2011a}).
  \textit{(a)}~Magnitude difference
  $\Delta H \equiv H_{\rm Photutils} - H_{\rm DAOPHOT}$ versus
  H-band magnitude; the red curve is the running median in a sliding
  window of 61~sources, and the dotted line marks the mean offset
  ($+0.036$~mag).
  \textit{(b)}~Distribution of $\Delta H$; the orange line indicates the
  median ($+0.033$~mag).
  The scatter is $\sigma = 0.074$~mag (MAD~$= 0.048$~mag);
  79~per cent of sources agree to $|\Delta H| < 0.1$~mag and
  98~per cent to $|\Delta H| < 0.2$~mag.
  The small systematic offset arises from background-estimation differences
  and is fully corrected by the photometric homogenisation
  (Section~\ref{sec:homog}).}
  \label{fig:validation}
\end{figure*}

The comparison is shown in Fig.~\ref{fig:validation}.
The mean offset is $\langle\Delta H\rangle = +0.036$~mag
(\Photutils$-$\DAOPHOT), with standard deviation 0.074~mag and median
absolute deviation MAD~$= 0.048$~mag.
The offset is positive, meaning \Photutils\ assigns slightly fainter
magnitudes than \DAOPHOT; this is consistent with the known difference in
background-estimation strategy (a global two-dimensional surface in
\Photutils\ versus a local per-source sky annulus in \DAOPHOT) and is
accounted for in the photometric homogenisation (Section~\ref{sec:homog}).
The agreement is quantitatively excellent: 79~per cent of matched sources
agree to within $|\Delta H| < 0.1$~mag, and 98~per cent to within
$|\Delta H| < 0.2$~mag.
The 10 most isolated bright stars agree to $-0.0004 \pm 0.0007$~mag,
confirming algorithmic equivalence for unblended, high-S/N sources.

The systematic offset of $+0.036$~mag is absorbed by the photometric
homogenisation procedure (Section~\ref{sec:homog}), which empirically
determines and applies inter-dataset corrections from the calibration-star
sample.
The residual cross-band systematic, if any, is estimated to be
$\lesssim 0.02$~mag, negligible relative to the AGB pulsation amplitudes
($A_K \sim 0.2$--2~mag) studied here.
We therefore adopt the \Photutils\ pipeline for the UHS K-band epoch.

%% ---- 2.5 Combined dataset ----
\subsection{Combined dataset and science context}
\label{sec:combined}

The four PSF catalogues (UFTI, UIST, WFCAM-PSF, UHS) span 15.1~yr
(HJD~2\,452\,919--2\,458\,438; 2003 October -- 2018 November).
For a Mira variable with $P = 500$~d, the UIST baseline alone samples
${\sim}2.8$ cycles; the full 15.1~yr baseline samples ${\sim}11$ cycles,
dramatically improving period accuracy and phase coverage.
For the most slowly pulsating variables ($P \sim 1500$~d) the combined
baseline provides ${\sim}3.7$ cycles, sufficient for reliable period
determination, whereas UIST alone would barely complete one cycle.
The 11.1-yr gap between the last WFCAM epoch (2007 October) and the UHS
epoch (2018 November) provides an independent constraint on secular
amplitude evolution inaccessible from the 2003--2007 data alone. A Monte
Carlo injection--recovery test confirms that the survey's non-uniform
cadence recovers genuine periodicities with ${>}95$~per cent reliability
across the 80--1500~d search range. Some structure in the empirical
$P_{\rm gls}$ distribution, most notably the well-known 1-yr sampling
alias common to ground-based monitoring, reflects the survey window
function rather than genuine variability and is treated with appropriate
caution.

\begin{table}
\centering
\caption{WFCAM epoch counts and temporal baseline per tile,
         underlying the new PSF reduction of
         Section~\ref{sec:wfcam_psf}. Individual observation
         dates and airmasses are given in table~1 of Paper~IV.}
\label{tab:wfcam_log}
\begin{tabular}{lcccc}
\hline
Tile & $N_K$ & HJD range ($-$2\,450\,000) & $N_J$ & $N_H$ \\
\hline
Q1 & 8 & 3632--4357 & 5 & 3 \\
Q2 & 8 & 3631--4362 & 2 & 2 \\
Q3 & 6 & 3631--4386 & 2 & 3 \\
Q4 & 5 & 3631--4397 & 1 & 2 \\
\hline
\end{tabular}
\end{table}

%%
%% ============================================================
%% SECTION 3 --- PHOTOMETRIC CATALOGUE CONSTRUCTION
%% ============================================================
%%
\section{Photometric catalogue construction}
\label{sec:photcat}

\subsection{Cross-matching and master catalogue assembly}
\label{sec:xmatch}

All four PSF catalogues were cross-matched against the UIST astrometric
reference frame using a 0.5-arcsec matching radius, chosen to be five times
the typical UIST centroiding precision (${\sim}0.1$~arcsec) and comfortably
larger than the UHS and WFCAM pixel scale (0.40~arcsec~pixel$^{-1}$).
The observed median match separation of 0.14~arcsec confirms that this
radius introduces no spurious matches.
The WFCAM aperture catalogue (Paper~IV) was excluded \DIFdelbegin \DIFdelend \DIFaddbegin \DIFadd{owing to blending biases in the crowded nuclear field }\DIFaddend (Section~\ref{sec:wfcam}).

The cross-matching yields a master catalogue of 847~unique stars
within the 3-arcmin-radius UFTI science field, each detected in at least
two of the four PSF datasets. The \DIFaddbegin \DIFadd{catalogue contains almost no stars in the
south-eastern part of the field (at offsets from the nucleus of more
than $36$~arcsec east and less than $30$--$40$~arcsec north;
${\approx}3.7$~arcmin$^2$), although Paper~I lists stars there. It
therefore covers ${\approx}12$~arcmin$^2$ (${\approx}0.9$~kpc$^2$ at
955~kpc) of the ${\approx}16$~arcmin$^2$ imaged by the UIST and UFTI
mosaics, and we exclude this region from the spatial analysis in
Section~\ref{sec:ampmag}.
The }\DIFaddend combined catalogue contains 21\,085~valid K-band measurements:
845~from UFTI (1 epoch per star; 845 of 847 stars covered\footnote{%
Master~IDs 520 and 808 have no UFTI detection; their absence is consistent
with proximity to a mosaic chip gap \DIFdelbegin \DIFdelend \DIFaddbegin \DIFadd{in }\DIFaddend the original UFTI reductions. Both
stars are detected in all three remaining PSF datasets and their absence
does not affect the science results.}),
14\,396~from UIST, 5\,536~from WFCAM-PSF, and 308~from UHS
(\DIFdelbegin \DIFdelend \DIFaddbegin \DIFadd{36}\DIFaddend ~per cent of the 847 stars).
All measurements are PSF-based: UFTI, UIST, and WFCAM-PSF via
\DAOPHOT/\ALLSTAR\ \citep{Stetson1987}; UHS via \Photutils\ v1.13
\citep{Bradley2023}.

The baseline spans HJD~2\,452\,918.83 to 2\,458\,438.50
(2003 October 5 to 2018 November 15; 5519.67~d $= 15.11$~yr).
Epoch counts per star range from 4 to 63 (mean 24.9, median 21);
the full epoch count distribution is given in
Table~\ref{tab:epochs} (Appendix~\ref{sec:appendix_epochs}).

\subsection{Photometric homogenisation}
\label{sec:homog}

\subsubsection{Reference system and motivation}

The four PSF catalogues were reduced independently, producing systematic
inter-dataset offsets that, if uncorrected, would masquerade as genuine
variability.
We adopt UFTI as the photometric reference, consistent with the 2MASS/Vega
K-band calibration chain established for Papers~I--V.
All offsets are $\Delta K = K_{\rm dataset} - K_{\rm UFTI}$; the applied
correction is $-\Delta K$.

\subsubsection{Calibration stars and threshold justification}

\DIFdelbegin \DIFdelend \DIFaddbegin \DIFadd{Low-scatter }\DIFaddend calibration stars were selected by requiring: (i)~detection
in all four PSF datasets; (ii)~$K < 17$~mag; and (iii)~RMS scatter
$\sRMS < 0.20$~mag over the 15-yr baseline.
The threshold $\sRMS < 0.20$~mag was determined by a sensitivity analysis
over the range 0.10--0.30~mag: \DIFdelbegin \DIFdelend \DIFaddbegin \DIFadd{of the 82~stars selected, fewer than 20 have
}\DIFaddend $\sRMS < 0.10$~mag \DIFdelbegin \DIFdelend (insufficient for robust offset estimation) \DIFdelbegin \DIFdelend \DIFaddbegin \DIFadd{and 45 have
}\DIFaddend $\sRMS < 0.15$~mag\DIFdelbegin \DIFdelend \DIFaddbegin \DIFadd{.
The derived offsets depend on this choice at the level of a few hundredths
of a magnitude: with only the }\DIFaddend 45~stars \DIFdelbegin \DIFdelend \DIFaddbegin \DIFadd{below 0.15~mag, the UIST and UHS
}\DIFaddend offsets change by \DIFdelbegin \DIFdelend \DIFaddbegin \DIFadd{0.01--0.02~mag and the WFCAM-PSF offset by 0.07~mag}\DIFaddend .
The threshold $\sRMS < 0.20$~mag was adopted as the best balance between
statistical robustness ($N = 82$) and contamination by low-level intrinsic
variables.

This yields 82 calibration stars spanning $K = 13.9$--17.0~mag,
with RMS values of 0.064--0.198~mag (mean 0.135~mag). \DIFaddbegin \DIFadd{Fewer than 20 of
them have RMS below 0.10~mag, the variability threshold of
Section~\ref{sec:var_detect}, so they are low-scatter stars common to
all four datasets rather than strictly non-variable stars.
}\DIFaddend To quantify the effectiveness of the correction, we define the
fractional change in scatter for each calibration star:

\begin{equation}
  \frac{\Delta\sigma_{\rm rms}}{\sigma_{\rm rms,before}} =
  \frac{\sigma_{\rm rms,after} - \sigma_{\rm rms,before}}{\sigma_{\rm rms,before}} \times 100,
  \label{eq:delta_rms}
\end{equation}

\noindent
where $\sigma_{\rm rms,before}$ and $\sigma_{\rm rms,after}$ are the
standard deviations of each star's raw and homogenised $K$-band light
curve, respectively, computed across all epochs at which that star was
detected; negative values indicate reduced scatter (improvement).

\begin{figure*}
\centering
  \includegraphics[width=\textwidth]
  {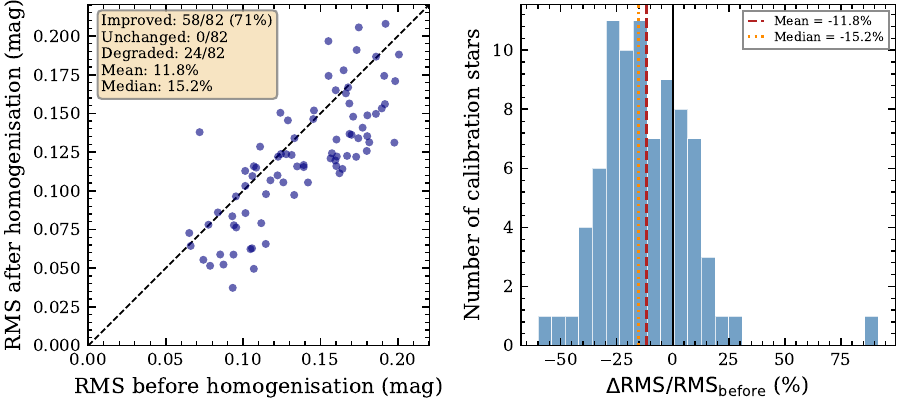}
  \caption{%
    Photometric homogenisation validation for the 82 \DIFdelbeginFL \DIFdelendFL \DIFaddbeginFL \DIFaddFL{low-scatter }\DIFaddendFL calibration
    stars (Section~\ref{sec:homog}).
    The left panel shows $\sigma_{\rm rms}$ per star\DIFaddbeginFL \DIFaddFL{, computed over all
    epochs of its combined light curve, }\DIFaddendFL before (horizontal axis) and after
    (vertical axis) applying the \DIFdelbeginFL \DIFdelendFL \DIFaddbeginFL \DIFaddFL{offsets of Table~\ref{tab:offsets}}\DIFaddendFL .
    Points below the dashed 1:1 line \DIFdelbeginFL \DIFdelendFL indicate reduced scatter
    post-correction; points above \DIFdelbeginFL \DIFdelendFL indicate increased scatter.
    The right panel shows the distribution of $\Delta\sigma_{\rm
    rms}/\sigma_{\rm rms,before}$ (Equation~\ref{eq:delta_rms}); negative
    values indicate reduced scatter (improvement) after homogenisation.
    The distribution has mean $-11.8$ and median $-15.2$~per cent (dashed
    and dotted vertical lines, respectively).
    Of 82 stars, 58 (\DIFdelbeginFL \DIFdelendFL \DIFaddbeginFL \DIFaddFL{71}\DIFaddendFL ~per cent) show improvement\DIFdelbeginFL \DIFdelendFL .
    The 24 stars whose RMS increases slightly \DIFdelbeginFL \DIFdelendFL \DIFaddbeginFL \DIFaddFL{span nearly }\DIFaddendFL the \DIFdelbeginFL \DIFdelendFL \DIFaddbeginFL \DIFaddFL{whole range
    of RMS (0.07--0.19~mag)}\DIFaddendFL ; a constant
    offset correction applied to a genuinely variable star may either
    increase or decrease the apparent RMS depending on the phase coverage
    of the observations relative to the pulsation cycle.
    The predominance of improvement, and the absence of a systematic
    increase for all 82 stars, confirms that the corrections are
    physically meaningful rather than an artefact of overfitting.
  }
  \label{fig:homog_val}
\end{figure*}

\subsubsection{Offsets and corrections}

For each non-reference dataset the systematic offset was estimated as the
median $\Delta K_i$ over 82~calibration stars; the median is robust against
outliers from blending or misidentification.
The formal uncertainty is
$\sigma_\delta \approx \sRMS/\sqrt{N} \approx 0.015$~mag, adopting the
calibration sample's mean $\sRMS = 0.135$~mag.
Results are in Table~\ref{tab:offsets}. \DIFaddbegin \DIFadd{UIST, the oldest dataset
(observations began in 2003), agrees with the UFTI reference to
$+0.0054$~mag, the smallest of the three offsets and within the
formal uncertainty; the age of the data therefore shows no sign of
affecting the calibration.
 }\DIFaddend 

\begin{table}
\centering
\caption{Photometric offsets relative to the UFTI reference system.}
\label{tab:offsets}
\begin{tabular}{lrrl}
\hline
Dataset   & $\Delta K_{\rm med}$ & Correction & Note \\
          & (mag) & (mag) & \\
\hline
UFTI      & $\phantom{+}0.0000$ & $\phantom{+}0.0000$ & Reference \\
UIST      & $+0.0054$           & $-0.0054$           & \\
WFCAM-PSF & $-0.0955$           & $+0.0955$           & \\
UHS       & $+0.2184$           & $-0.2184$           & $^{a}$ \\
\hline
\end{tabular}
\begin{minipage}{\linewidth}
\smallskip
\textit{Notes.} Formal uncertainty $\approx 0.015$~mag per offset.
The correction is exact: $K_{\rm corr} \equiv K_{\rm orig} + \delta$
(residuals $= 0.000000$~mag for all 21\,085 measurements).
($a$)~The large UHS offset reflects the difference between the arbitrary
\Photutils\ zero-point ($\mathrm{ZP}_{\rm inst} = 25.0$~mag) and the
2MASS-calibrated UFTI system \citep{Hodgkin2009}; since the offset is
determined empirically from 82~\DIFdelbeginFL \DIFdelendFL \DIFaddbeginFL \DIFaddFL{low-scatter }\DIFaddendFL calibration stars and applied
to machine precision, its absolute value does not affect any science
result.
The \DIFdelbeginFL \DIFdelendFL \DIFaddbeginFL \DIFaddFL{slopes }\DIFaddendFL of $\Delta K$ \DIFdelbeginFL \DIFdelendFL \DIFaddbeginFL \DIFaddFL{against the catalogue mean }\DIFaddendFL $K$ \DIFdelbeginFL \DIFdelendFL \DIFaddbeginFL \DIFaddFL{are
$-0.01\pm0.02$, $-0.04\pm0.04$ and $-0.04\pm0.04$}\DIFaddendFL ~mag~mag$^{-1}$ \DIFdelbeginFL \DIFdelendFL \DIFaddbeginFL \DIFaddFL{for UIST,
WFCAM-PSF and UHS, none of them significant }\DIFaddendFL for $N = 82$\DIFdelbeginFL \DIFdelendFL \DIFaddbeginFL \DIFaddFL{, and a }\DIFaddendFL constant
correction is adopted\DIFdelbeginFL \DIFdelendFL \DIFaddbeginFL \DIFaddFL{. A slope of 0.04~mag~mag$^{-1}$ amounts to about
0.1~mag between $K = 14$ and 17~mag, small compared with the
amplitudes of 90~per cent of the variables ($A_K \geq 0.5$}\DIFaddendFL ~mag).
\end{minipage}
\end{table}

\subsubsection{Validation of homogenisation}

Of the 82~calibration stars, 58 (\DIFdelbegin \DIFdelend \DIFaddbegin \DIFadd{71}\DIFaddend ~per cent) show reduced RMS scatter
after correction\DIFdelbegin \DIFdelend \DIFaddbegin \DIFadd{; the mean change over all 82 stars is $-11.8$}\DIFaddend ~per cent
(\DIFaddbegin \DIFadd{median $-15.2$~per cent; }\DIFaddend Fig.~\ref{fig:homog_val}).
The remaining 24~stars (\DIFdelbegin \DIFdelend \DIFaddbegin \DIFadd{29}\DIFaddend ~per cent) show slight increases: these have
low-level intrinsic variability (\DIFdelbegin \DIFdelend \DIFaddbegin \DIFadd{their RMS before correction, 0.07--0.19~mag,
spans nearly the whole range of the sample}\DIFaddend ), and a constant offset
correction can either increase or decrease apparent scatter depending on
phase coverage.
The mixed outcome is expected \DIFdelbegin \DIFdelend \DIFaddbegin \DIFadd{because the calibration stars are not strictly
constant}\DIFaddend , and argues against overfitting.

The \DIFdelbegin \DIFdelend \DIFaddbegin \DIFadd{offsets show no significant dependence on magnitude
(}\DIFaddend Table~\ref{tab:offsets} note~$a$)\DIFdelbegin \DIFdelend \DIFaddbegin \DIFadd{, consistent with a }\DIFaddend zero-point shift
rather than a systematic error in the PSF fitting.
The homogenised catalogue (hereafter the \textit{Paper~VII catalogue})
comprises 21\,085 K-band measurements of 847~stars spanning 15.11~yr
and is available at the CDS.

%% ============================================================
%% SECTION 4 --- VARIABLE STAR IDENTIFICATION AND CLASSIFICATION
%% ============================================================

\section{Variable star identification and classification}
\label{sec:variables}

\subsection{Variability detection}
\label{sec:var_detect}

We identified variable stars using a robust root-mean-square
(RMS) selection criterion, widely employed in large-scale
long-period variable (LPV) surveys
\citep{Soszynski2009, Wood1999, Yuan2018}.
For each star we computed the standard deviation of the
homogenised K-band magnitudes:
\begin{equation}
  \sigma_{\mathrm{K}} =
  \sqrt{\frac{\displaystyle\sum_{i=1}^{N}
  \bigl(K_i - \bar{K}\bigr)^2}{N - 1}},
  \label{eq:sigma_k}
\end{equation}
where $K_i$ are the individual epoch measurements,
$\bar{K}$ is the unweighted mean magnitude, and $N$ is the
number of valid epochs per star.

We adopt $\sigma_{\mathrm{K}}$ in preference to the Stetson
\citeyearpar{Stetson1996} $L$-index used in Papers~I and~IV for
two reasons.
First, the $L$-index requires contemporaneous multi-band
observations to form correlated residual pairs; in our dataset,
$J$- and $H$-band epochs are available for only a subset of
UIST and WFCAM nights and are entirely absent from the UFTI and
UHS datasets.
Second, and more fundamentally, $\sigma_{\mathrm{K}}$ is directly
comparable across all four heterogeneous datasets after homogenisation,
whereas variability indices calibrated on the UIST cadence (23 epochs,
2003--2007) would not transfer consistently to the single-epoch UFTI
and UHS datasets.
The single-band $\sigma_{\mathrm{K}}$ criterion makes full use of
all 21\,085 K-band measurements and is the standard variability
estimator for single-band multi-epoch LPV surveys of this type
\citep{Soszynski2009, Yuan2018}.

A star is classified as variable if
$\sigma_{\mathrm{K}} > 0.100$\,mag.
This threshold corresponds to approximately $5\sigma$ above
our median photometric precision of $\sim$0.02\,mag for
$K < 17$\,mag, ensuring high statistical confidence in
variability detection while minimising contamination from
photometric noise.
The threshold is motivated by extensive precedent in LPV
surveys \citep{Soszynski2009}\DIFdelbegin \DIFdelend \DIFaddbegin \DIFadd{. Non-variable }\DIFaddend stars span
$\sigma_{\mathrm{K}} = 0.033$--$0.100$\,mag \DIFdelbegin \DIFdelend and variable stars
\DIFdelbegin \DIFdelend $0.101$--$1.951$\,mag \DIFdelbegin \DIFdelend \DIFaddbegin \DIFadd{(Fig.~\ref{fig:rms_dist})}\DIFaddend .

Of 847 stars in the homogenised catalogue, 771 (\DIFdelbegin \DIFdelend \DIFaddbegin \DIFadd{91 }\DIFaddend per cent)
are significantly variable under this criterion -- markedly higher
than the variability fractions reported for the same central region
by Paper~I (812/18\,398 UIST detections, 4.4~per cent) and Paper~IV
(667/11\,114 WFCAM detections at $L>1.5$, 6.0~per cent).
This high variability fraction reflects the magnitude-limited
and AGB-dominated nature of the Paper~VII catalogue, which
targets stars pre-selected by brightness and multi-dataset
detection rather than surveying the full stellar population;
the 15.1-yr baseline is specifically sensitive to semi-periodic
variability on timescales $P \sim 1000$--5000~d that would
appear non-variable in surveys with baselines of 1--3~yr
\citep{Wood1999}. Timescales of this order need not indicate
fundamental-mode radial pulsation: comparably long periods have
been attributed to binarity-induced long secondary periods in
red supergiants (e.g.\ $P\approx2000$~d in Betelgeuse), as well
as to genuine pulsation in the most massive, dust-enshrouded AGB
stars with OH masers, where periods of order 1200~d are observed
\citep{Wood1992, Goldman2017}.

\subsection{Amplitude measurement and classification}
\label{sec:amplitude}

We characterise the pulsation amplitude of each variable
star by its peak-to-peak magnitude range:
\begin{equation}
  A_{\mathrm{K}} = K_{\mathrm{max}} - K_{\mathrm{min}},
  \label{eq:amplitude}
\end{equation}
where $K_{\mathrm{max}}$ and $K_{\mathrm{min}}$ are the
faintest and brightest magnitudes recorded in the homogenised
catalogue, respectively.
This estimator is standard in near-infrared LPV surveys
\citep{Whitelock2008, Soszynski2009}.

We adopt peak-to-peak amplitude rather than a sinusoidal
amplitude derived from period fitting for two reasons.
First, this is a catalogue paper: period determination is
deferred to Paper~VIII, and a period-fitted sinusoidal
amplitude would presuppose a reliable period already in
hand, which cannot be guaranteed prior to that vetting.
Second, $A_K$ is a purely observational, model-independent
quantity computable for all 771 variable stars regardless
of whether their light curves are periodic, semi-regular,
or irregular.
We note two limitations inherent to this estimator.
For the 17 stars with $N_{\rm ep} \leq 10$, $A_K$ is a
lower limit because the true photometric extrema may not
have been sampled.
For 21 borderline variable stars, $\sigma_{\mathrm{K}}$
exceeds the 0.10~mag threshold owing to a single deviant
epoch; inspection of their photometric uncertainties shows
that in 16 of 21 cases the outlier epoch has an uncertainty
exceeding the star's median by a factor $\geq 1.5$,
identifying it as a likely bad measurement
(see Appendix~\ref{sec:appendix_borderline} for a complete
table and analysis).
These stars are retained in the catalogue and flagged
\texttt{QF~=~1} in the CDS release.
For the \DIFdelbegin \DIFdelend \DIFaddbegin \DIFadd{770 variables with $N_{\rm ep} \geq 5$, a preliminary
uncertainty-weighted Generalised Lomb--Scargle fit (trial periods
80--1500~d) gives a median ratio $A_{\rm ptp}/A_K^{\rm GLS} = 1.8$
(16th--84th percentiles 1.2--3.4), where $A_K^{\rm GLS}$ is the full
amplitude of the best-fitting sinusoid. The }\DIFaddend observed peak-to-peak
amplitude \DIFdelbegin \DIFdelend \DIFaddbegin \DIFadd{exceeds this fitted amplitude for 95~per~cent of these
stars, as expected if a single sinusoid captures only the coherent
periodic component, whereas $A_{\rm ptp}$ also includes cycle-to-cycle
changes, irregular variability and photometric scatter.
Both $A_{\rm ptp}$ and this preliminary $A_K^{\rm GLS}$ are
included as separate columns in the CDS catalogue release,
allowing direct comparison with the rigorous period-fitted
amplitudes and classification to be presented in Paper~VIII.
The two amplitude measures are well correlated (Spearman rank
coefficient 0.74); when the class boundaries defined below are
rescaled by the median }\DIFaddend $A_{\rm ptp}/A_K^{\rm GLS}$ \DIFdelbegin \DIFdelend \DIFaddbegin \DIFadd{ratio,
64~per~cent of the variables retain their class (45~per~cent
without rescaling). The amplitude classes adopted here are
therefore descriptive, and the classification of individual stars
may change once periods are determined }\DIFaddend in Paper~VIII.

Following these precedents, we divide the 771 variable stars
into three amplitude classes (Table~\ref{tab:variables}):

\begin{itemize}

  \item \textbf{High-amplitude variables}
    ($A_{\mathrm{K}} \geq 1.0$\,mag; 385 stars,
    \DIFdelbegin \DIFdelend \DIFaddbegin \DIFadd{45 }\DIFaddend per cent of the full sample).
    The mean scatter for this class is
    $\langle\sigma_{\mathrm{K}}\rangle = 0.525$\,mag,
    exceeding photometric precision by a factor of $\sim$26.
    Their large, quasi-regular pulsation amplitudes are
    characteristic of Mira variables on the thermally
    pulsing AGB \citep{Wood1999}; they are the primary
    targets for the period-luminosity analysis in Paper~VIII.

  \item \textbf{Moderate-amplitude variables}
    ($0.5 \leq A_{\mathrm{K}} < 1.0$\,mag; 308 stars,
    \DIFdelbegin \DIFdelend \DIFaddbegin \DIFadd{36 }\DIFaddend per cent).
    With $\langle\sigma_{\mathrm{K}}\rangle = 0.204$\,mag
    ($\sim$10 times the photometric precision), this
    population is consistent with semi-regular AGB variables
    exhibiting less coherent pulsation than Miras.

  \item \textbf{Low-amplitude variables}
    ($A_{\mathrm{K}} < 0.5$\,mag; 78 stars,
    \DIFdelbegin \DIFdelend \DIFaddbegin \DIFadd{9 }\DIFaddend per cent).
    These marginally variable sources have
    $\langle\sigma_{\mathrm{K}}\rangle = 0.134$\,mag.
    They may include small-amplitude irregular variables (such as
red supergiants) or stars at the onset of the thermally pulsing
AGB phase.

\end{itemize}

\noindent
The remaining 76 stars (\DIFdelbegin \DIFdelend \DIFaddbegin \DIFadd{9 }\DIFaddend per cent) show no significant
variability ($\sigma_{\mathrm{K}} \leq 0.100$\,mag)\DIFdelbegin \DIFdelend .

\subsection{Variability characteristics}
\label{sec:var_char}

\begin{figure}
  \includegraphics[width=\columnwidth]{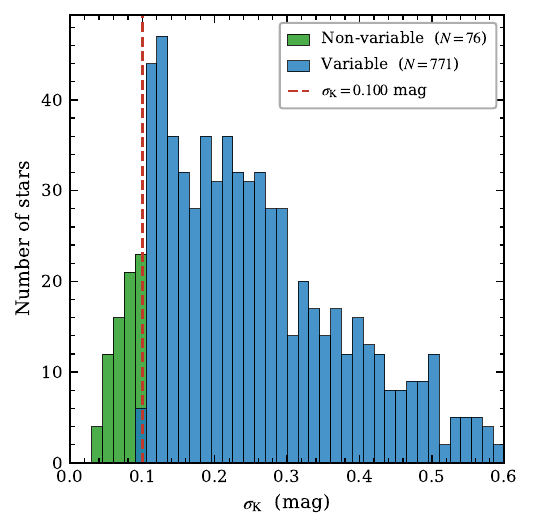}
  \caption{%
    Distribution of $K$-band RMS scatter $\sigma_K$
    (equation~\ref{eq:sigma_k}) for all 847~stars in the
    homogenised catalogue, binned at $\Delta\sigma_K = 0.01$~mag.
    Non-variable stars ($\sigma_K \leq 0.100$~mag; green,
    $N = 76$) and variable stars ($\sigma_K > 0.100$~mag; blue,
    $N = 771$) are cleanly separated at the adopted threshold
    (red dashed line), which lies $\approx\!5\sigma$ above the
    typical photometric precision of $\sim 0.02$~mag for
    $K < 17$~mag.
    The number of stars per bin increases by roughly 50~per cent
immediately above the threshold, consistent with the onset of
the intrinsic-scatter-dominated (variable) population.
  }
  \label{fig:rms_dist}
\end{figure}

\begin{figure*}
  \centering
  \includegraphics[width=\textwidth]{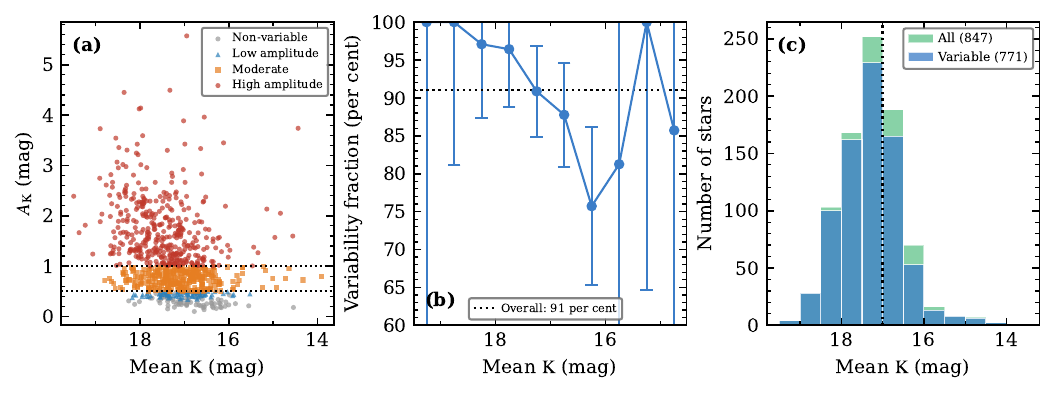}
  \caption{%
    Variability statistics as a function of mean $K$-band magnitude
    for the 847-star Paper~VII catalogue.
    (a)~Peak-to-peak amplitude $A_K = K_{\rm max}-K_{\rm min}$
    (equation~\ref{eq:amplitude}), colour-coded by amplitude class
    (Section~\ref{sec:amplitude}): high-amplitude (red; $A_K\geq1.0$~mag,
    $N=385$), moderate (orange; $0.5\leq A_K<1.0$~mag, $N=308$),
    low-amplitude (blue; $A_K<0.5$~mag, $N=78$), and non-variable
    (grey; $N=76$); dotted lines mark the class boundaries.
        (b)~Variability fraction in 0.5-mag bins, with $1\sigma$ Poisson
    uncertainties; the dotted line marks the overall fraction of
    \DIFdelbeginFL \DIFdelendFL \DIFaddbeginFL \DIFaddFL{91}\DIFaddendFL ~per cent (771/847).
    \DIFdelbeginFL \DIFdelendFL (c)~$K$-band luminosity function: all 847~stars (green) and the
    771~classified variables (blue), binned at $\Delta K=0.5$~mag; the
    dotted line marks $K=17$~mag, the thermally pulsing AGB tip at the
    distance of M33 ($\mu=24.9$~mag; \citealt{Javadi2011a}).
  \DIFdelbeginFL \DIFdelendFL }
  \label{fig:amp_var_lf}
\end{figure*}

The $\sigma_{\mathrm{K}}$ distribution
(Fig.~\ref{fig:rms_dist}) \DIFdelbegin \DIFdelend \DIFaddbegin \DIFadd{has }\DIFaddend star counts increasing
steadily through the adopted threshold at $\sigma_{\mathrm{K}}
= 0.100$\,mag\DIFdelbegin \DIFdelend \DIFaddbegin \DIFadd{, with no }\DIFaddend discrete gap between \DIFdelbegin \DIFdelend \DIFaddbegin \DIFadd{the two populations}\DIFaddend .
Fig.~\ref{fig:amp_var_lf} places this threshold in the context
of the full sample.
The amplitude--magnitude diagram (panel~a) reveals that
high-amplitude sources (Mira candidates) are present across
the full surveyed magnitude range ($13.9 \lesssim K \lesssim
19.5$\,mag)\DIFdelbegin \DIFdelend .
The variability fraction (panel~b) \DIFdelbegin \DIFdelend \DIFaddbegin \DIFadd{lies between 76 and
100}\DIFaddend ~per cent \DIFdelbegin \DIFdelend \DIFaddbegin \DIFadd{in }\DIFaddend all magnitude bins, \DIFdelbegin \DIFdelend \DIFaddbegin \DIFadd{rising from 76~per cent at
$K=16.0$--$16.5$~mag to 96~per cent or more at $K>17.5$~mag
(91~per cent overall)}\DIFaddend ; the corresponding luminosity function
(panel~c) is discussed in Section~\ref{sec:lf}.

\begin{figure}
  \includegraphics[width=\textwidth, height=0.25\textheight, 
keepaspectratio]{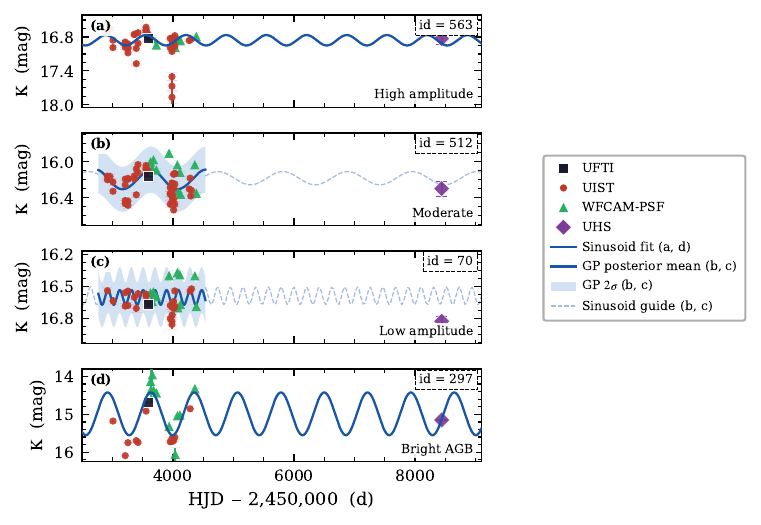}
  \caption{%
    Representative $K$-band light curves spanning the full
    amplitude range of the catalogue\DIFdelbeginFL \DIFdelendFL , \DIFdelbeginFL \DIFdelendFL with $1\sigma$ error bars\DIFdelbeginFL \DIFdelendFL .
    The blue curve shows an illustrative fit to guide the eye:
for panels~(b) and~(c) this is the posterior mean of a
semi-parametric Gaussian Process (GP) model
\citep{He2016, Yuan2018}, with the shaded band giving its
$2\sigma$ interval over the well-sampled baseline; for panels~(a) and~(d), where the GP fit did not converge
robustly, it is instead a simple sinusoid at the GLS
period. Rigorous period determination is presented in
Paper~VIII.
    \DIFdelbeginFL \DIFdelendFL (a)~High-amplitude Mira candidate
        (id\,=\,563; $A_K = 2.39$\,mag, $N = 51$,
        $P_{\rm fit} \approx 664$\,d).
    (b)~Moderate-amplitude semi-regular variable
        (id\,=\,512; $A_K = 0.63$\,mag, $N = 60$,
        $P_{\rm GP} \approx 1069$\,d).
    (c)~Low-amplitude variable
        (id\,=\,70; $A_K = 0.49$\,mag, $N = 37$,
        $P_{\rm GP} \approx 210$\,d).
    (d)~Luminous AGB variable at the bright end of the
        sample (id\,=\,297; $A_K = 2.13$\,mag,
        $\langle K\rangle = 14.81$\,mag, $N = 24$,
        $P_{\rm fit} \approx 716$\,d).
  }
  \label{fig:lightcurves}
\end{figure}

Representative K-band light curves spanning the full range of
variability amplitude are shown in Fig.~\ref{fig:lightcurves}.
The four panels illustrate the diversity of light-curve morphology
among the variable star population of the M33 central kiloparsec,
drawn from the three amplitude classes defined in
Section~\ref{sec:variables}: high amplitude ($A_K \geq 1.0$\,mag),
moderate ($0.5 \leq A_K < 1.0$\,mag), and low amplitude
($A_K < 0.5$\,mag, $\sigma_{\rm rms} > 0.10$\,mag).
A fourth panel illustrates a bright AGB variable
($\langle K\rangle < 15$\,mag) representative of the
dust-enshrouded population identified in
Section~\ref{sec:spitzer_xmatch}.
Each panel combines photometry from all four UKIRT instruments
after cross-calibration (Section~\ref{sec:homog}),
spanning the full 15.11-yr baseline.
\DIFdelbegin \DIFdelend 

To guide the eye, we overlay two complementary fits. For
panels~(b) and~(c), a semi-parametric Gaussian process (GP)
model \citep{He2016, Yuan2018} decomposes the variability into
a coherent periodic component, capturing the dominant pulsation
period, and a stochastic component accounting for the
cycle-to-cycle variability intrinsic to semi-regular variables
\citep{TempletonKarovska2009}; the blue line and shaded band
show the GP posterior mean and $2\sigma$ predictive interval,
plotted over the well-sampled baseline only, since extending
the band across the 2008--2017 observing gap would imply an
unwarranted degree of confidence. For panels~(a) and~(d), the
GP fit did not converge on a stable periodic solution; these
two stars are instead shown with a simple sinusoid at the GLS
period, plotted continuously for visual comparison. For visual
continuity outside the GP-covered baseline, panels~(b) and~(c)
additionally show a faint dashed sinusoid at the same GLS
period; given its poor fit quality there, this is a visual
guide only and carries no statistical weight. We emphasise that
all overlaid curves are illustrative aids to visualise the
dominant periodicity; quantitative period determination is
presented in Paper~VIII.

Panel~(a) shows a classical high-amplitude variable
(Star~563, $A_K = 2.39$\,mag, $N = 51$, $P \approx 665$\,d),
whose coherent sinusoidal modulation is characteristic of an
oxygen-rich Mira pulsating on the fundamental mode.
Panel~(b) shows a moderate-amplitude variable
(Star~512, $A_K = 0.63$\,mag, $N = 60$, $P \approx 1069$\,d);
its long inferred period places it among the most evolved
AGB stars in the sample.
Panel~(c) shows a low-amplitude semi-regular variable
(Star~70, $A_K = 0.49$\,mag, $N = 37$, $P \approx 210$\,d),
below the conventional Mira amplitude threshold of
$A_K = 1.0$\,mag.
With $\approx\!26$ observed pulsation cycles over the
15-yr baseline, the period of this star is among the
best-constrained in the catalogue.
Panel~(d) shows a luminous AGB variable near the bright
end of the sample
(Star~297, $A_K = 2.13$\,mag,
$\langle K\rangle = 14.81$\,mag, $N = 24$,
$P \approx 714$\,d), with $M_K \approx -10.1$\,mag at the
M33 distance modulus ($\mu = 24.9$\,mag;
\citealt{Javadi2011a}).
Such luminous, large-amplitude variables are prime candidates
for dust-enshrouded AGB stars, consistent with the
$K - [3.6]$ excess observed in the Spitzer cross-match
(Section~\ref{sec:spitzer_xmatch}).

Across all four panels, photometry from the four instruments
lies on a consistent photometric scale without visible
systematic offsets, directly validating the cross-calibration
of Section~\ref{sec:homog}.
The single UHS epoch (rightmost symbol in each panel,
$\mathrm{HJD} - 2\,450\,000 = 8438$)
extends the temporal baseline by $\approx\!11$\,yr beyond
the last WFCAM observation, enabling reliable period
determination for variables with $P \gtrsim 400$\,d.

%% -----------------------------------------------------------------
%% TABLE --- Variable star classification
%% -----------------------------------------------------------------
\begin{table}
  \caption{Variable star classification summary}
  \label{tab:variables}
  \begin{tabular}{lcccc}
  \hline
  Class & Criterion & $N$ &
    $\langle\sigma_{\mathrm{K}}\rangle$ &
    $\langle A_{\mathrm{K}}\rangle$ \\
  & & & (mag) & (mag) \\
  \hline
  High amplitude
    & $A_{\mathrm{K}} \geq 1.0$       & 385 & 0.525 & \DIFdelbeginFL \DIFdelendFL \DIFaddbeginFL \DIFaddFL{1.77 }\DIFaddendFL \\
  Moderate
    & $0.5 \leq A_{\mathrm{K}} < 1.0$ & 308 & 0.204 & \DIFdelbeginFL \DIFdelendFL \DIFaddbeginFL \DIFaddFL{0.75 }\DIFaddendFL \\
  Low amplitude
    & $A_{\mathrm{K}} < 0.5$          &  78 & \DIFdelbeginFL \DIFdelendFL \DIFaddbeginFL \DIFaddFL{0.121 }\DIFaddendFL & \DIFdelbeginFL \DIFdelendFL \DIFaddbeginFL \DIFaddFL{0.44 }\DIFaddendFL \\
  Non-variable
    & $\sigma_{\mathrm{K}} \leq 0.100$&  76 & 0.076 & \DIFdelbeginFL \DIFdelendFL \DIFaddbeginFL \DIFaddFL{0.29 }\DIFaddendFL \\
  \hline
  All stars & & 847 & \DIFdelbeginFL \DIFdelendFL \DIFaddbeginFL \DIFaddFL{0.331 }\DIFaddendFL & 1.14 \\
  \hline
  \end{tabular}
  \begin{minipage}{\columnwidth}
  \smallskip
  \textit{Notes.} $\sigma_{\mathrm{K}}$ = standard deviation of homogenised
  $K$-band magnitudes (equation~\ref{eq:sigma_k}).
  $A_{\mathrm{K}} = K_{\mathrm{max}} - K_{\mathrm{min}}$ = peak-to-peak
  amplitude (equation~\ref{eq:amplitude}).
  Angle brackets denote class means.
  \DIFdelbeginFL \DIFdelendFL \end{minipage}
\end{table}

%% -----------------------------------------------------------------
%% FIGURE CAPTIONS
%% -----------------------------------------------------------------

%% ============================================================
%% SECTION 5 --- CATALOGUE PROPERTIES
%% ============================================================
\section{Catalogue properties}
\label{sec:results}

\subsection{Luminosity function}
\label{sec:lf}

The K-band luminosity function (LF) of the 847-star catalogue
(panel~c of Fig.~\ref{fig:amp_var_lf}, introduced in
Section~\ref{sec:var_char}) traces the evolved stellar population of the M33 central
kiloparsec over the range $K = 13.9$--19.5~mag
($M_K \approx -11.0$ to $-5.4$~mag at $\mu = 24.9$~mag;
\citealt{Javadi2011a}).
The bright end ($K < 15$~mag; $M_K < -9.9$~mag) is dominated by the most
luminous AGB stars and RSGs\DIFdelbegin \DIFdelend \DIFaddbegin \DIFadd{: nine of the ten stars in this range are
variable, four of them high-amplitude and five of moderate amplitude.
Among the 292 }\DIFaddend stars brighter than $K = 17$~mag, \DIFdelbegin \DIFdelend \DIFaddbegin \DIFadd{248 (85}\DIFaddend ~per cent\DIFdelbegin \DIFdelend \DIFaddbegin \DIFadd{) are
variable and 79 are }\DIFaddend high-amplitude variables, consistent with
thermally pulsing AGB evolution \citep{Whitelock2008, Ita2004a}.
The LF peaks near $K \approx 17$~mag ($M_K \approx -7.9$~mag),
corresponding to the thermally pulsing AGB tip and the brightest population
of oxygen-rich Mira variables \citep{Feast1989, Whitelock2008}.
Below this, the rising RGB population contributes increasingly, though the
requirement of multi-dataset detection limits completeness at
$K \gtrsim 18$~mag.

All 847~stars in the Paper~VII catalogue have UIST counterparts in
Paper~I \citep{Javadi2011a}\DIFdelbegin \DIFdelend \DIFaddbegin \DIFadd{. The homogenisation reduces the RMS
scatter of the 82 calibration stars by a mean of 11.8}\DIFaddend ~per cent
\DIFdelbegin \DIFdelend \DIFaddbegin \DIFadd{(Section~\ref{sec:homog})}\DIFaddend .

\subsection{Amplitude--magnitude relation}
\label{sec:ampmag}

The amplitude--magnitude relation (panel~a of Fig.~\ref{fig:amp_var_lf},
introduced in Section~\ref{sec:var_char}) reveals the
expected sequence for evolved pulsating stars: amplitudes decrease toward
brighter magnitudes, a consequence of the logarithmic magnitude scale for
a fixed fractional change in luminosity \citep{vanLoon2008}.
The brightest stars ($K < 15$~mag) span the full amplitude range,
including several with $A_K < 1.0$~mag alongside genuine high-amplitude
Miras, rather than showing exclusively high amplitudes.
This inverse trend is consistent with the period-luminosity and
period-amplitude relations of Mira variables in the LMC
\citep{Feast1989, Ita2004a}, and the high-amplitude tail of our sample
overall remains dominated by classical oxygen-rich Miras rather than
semi-regular or irregular variables. \DIFaddbegin \DIFadd{For comparison, Galactic Miras
in the bulge and disc \citep{Nikzat2022} and in the nuclear stellar
disc \citep{Sanders2022, Sanders2024} are distinguished by $K_s$
amplitudes above ${\approx}0.4$ and ${\approx}0.25$--0.4~mag, respectively,
below the lower boundary of our moderate-amplitude class. These
amplitudes are not necessarily defined in the same way as our
peak-to-peak values, so a quantitative comparison will be possible
once periods are determined for the M33 sample in Paper~VIII.
}

\begin{figure*}
  \includegraphics[width=\textwidth]{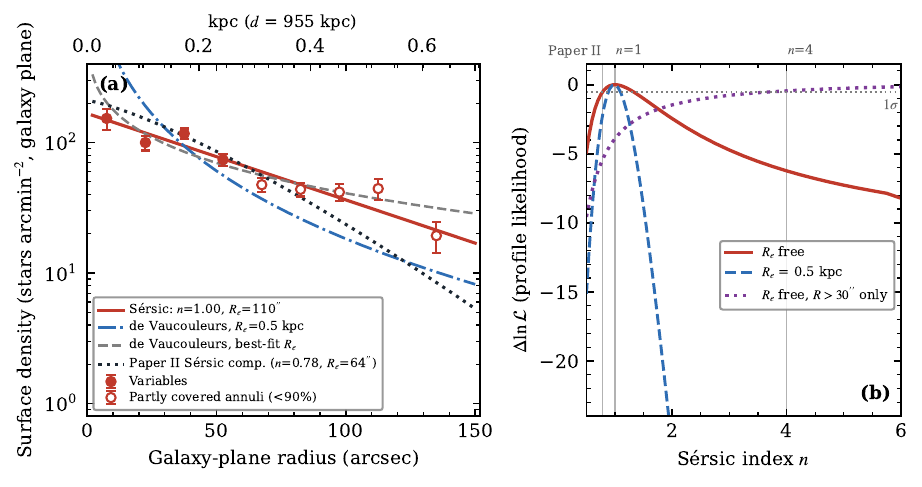}
  \caption{\DIFaddFL{Radial distribution of the variables in the galaxy plane
    (position angle $23\degr$, inclination $56\degr$; Paper~II),
    within $r<84$~arcsec of the nucleus and excluding the
    south-eastern part of the field that contains almost no catalogued stars
    (Section~\ref{sec:xmatch}). \textit{(a)}~Surface density (open
    symbols: annuli less than 90~per~cent inside the analysed
    region) with maximum-likelihood profiles: the best-fitting
    S\'ersic profile, de~Vaucouleurs ($n=4$) profiles with
    $R_e=0.5$~kpc and with the best-fitting $R_e$, and the shape of the
    Paper~II S\'ersic component (all with fitted amplitude). The top
    axis assumes 955~kpc ($\mu=24.9$~mag). \textit{(b)}~Profile
    likelihood of the S\'ersic index for free $R_e$, for
    $R_e=0.5$~kpc, and for free $R_e$ excluding $R<30$~arcsec.}}
  \label{fig:radial_profile}
\end{figure*}

\DIFadd{To assess whether the LPV population traces a bulge-like spatial
distribution, such as the central bulge proposed by \citet{Minniti1993},
we repeated the de-projected S\'ersic-profile analysis of Paper~II
(same nucleus position, $\mathrm{PA}=23\degr$, $i=56\degr$) on the 771
variables (Fig.~\ref{fig:radial_profile}). We fitted the unbinned
positions by maximum likelihood within $r<84$~arcsec of the nucleus,
excluding the south-eastern part of the field that contains almost no
catalogued stars (Section~\ref{sec:xmatch}; 538~variables remain). We
find $n=1.0^{+0.3}_{-0.2}$ and $R_e=110^{+39}_{-19}$~arcsec
(0.51~kpc at 955~kpc; profile-likelihood uncertainties), i.e. a
near-exponential distribution, with an index consistent with that of
the AGB stars in Paper~II ($n=0.78$) and an effective radius close to
the 0.5~kpc found for the bulge by \citet{Minniti1993}. A
de~Vaucouleurs profile ($n=4$) with $R_e=0.5$~kpc is rejected
($\Delta\ln\mathcal{L}=-108$ relative to the best fit); with $R_e$ free
it is also disfavoured ($\Delta\ln\mathcal{L}=-6$) and would require
$R_e\approx13$~kpc, an almost flat distribution over our field. We
therefore find no evidence that the variables follow a classical
bulge. The fitted $n$ and $R_e$ change by up to ${\approx}0.5$ and
${\approx}0.3$~kpc for other plausible geometries and for fitting
radii of 72--108~arcsec (within 60~arcsec $R_e$ is only bounded from
below), and the conclusion depends on the completeness in the crowded
inner region: normalised on the stars outside $R=30$~arcsec, the $n=4$,
$R_e=0.5$~kpc profile predicts ${\approx}340$ variables inside it,
against 89 observed. The best-fitting profile describes the counts in
annuli only approximately ($\chi^2=17.0$ for 9 annuli and about 6
degrees of freedom). We regard these results as a consistency check on
Paper~II rather than an independent structural measurement.
}\DIFaddend 

The moderate-amplitude group ($0.5 \le A_K < 1.0$~mag; 308 stars) contains
a mixture of semi-regular variables and long-period irregulars; precise
classification requires period analysis (Paper~VIII).
Stars with $A_K < 0.5$~mag and $\sRMS > 0.10$~mag (78~stars) populate
the low-amplitude regime consistent with semi-regular variables and OGLE
Small Amplitude Red Giants \citep[OSARGs;][]{Soszynski2004}; period
determination from the 15-yr baseline will establish their precise
classification in Paper~VIII.

\subsection{Multi-wavelength analysis}
\label{sec:multiwave}
Near-infrared photometry alone cannot distinguish between AGB stars
with differing degrees of circumstellar dust production, since
$K$-band amplitude and mean magnitude are influenced by
evolutionary state, metallicity, and pulsation mode in addition to
mass-loss history.
Mid-infrared photometry provides a complementary, more direct
diagnostic: dust preferentially absorbs and re-emits starlight at
these wavelengths, so a star's $K-[3.6]$ colour excess traces its
circumstellar dust content largely independently of the
near-infrared variability properties already characterised in
Section~\ref{sec:var_char}.
Cross-matching our catalogue with mid-infrared photometry therefore
allows us to identify which of our variable stars are the most
dust-productive, and hence most evolved, members of the AGB
population.

\subsubsection{Search for optimal mid-infrared photometric data}
\label{sec:jwst_search}

To characterise the circumstellar dust content of our variable star
sample, mid-infrared photometry at $[3.6]$ and $[4.5]\,\micron$ is
required for the $\langle\mathrm{K}\rangle-[3.6]$ colour excess
diagnostic. We first assessed whether \textit{James Webb Space
Telescope} (\textit{JWST}) data were available for the central
$3\,\mathrm{arcmin}$ of~M33.

The PHANGS-\textit{JWST} Cycle~1 Treasury (GO-2107;
\citealt{Lee2023}) does not include M33; at \DIFdelbegin \DIFdelend \DIFaddbegin \DIFadd{${\approx}955\,\mathrm{kpc}$
}\DIFaddend it is too nearby for the PHANGS strategy, which targets galaxies at
$D=5$--$20\,\mathrm{Mpc}$ \citep{Williams2024}.
Programme GO-2128 (\citealt{Peltonen2024}) observed M33 with MIRI in
F560W and F2100W, but the field centre lies
${\approx}28\,\mathrm{arcmin}$ south-west of the nucleus, covering
the southern spiral arm with no overlap with our field.
NIRCam parallels \citep{Sarbadhicary2025} are co-spatial with the
MIRI field and targeted supernova remnant science, not AGB photometry.
Programme GO-3436 (PI: E.~Rosolowsky; Cycle~2) obtained NIRCam
(F150W, F335M) and MIRI (F770W) imaging of the M33 central field in
2024 January, targeting interstellar medium and supernova remnant
science; no AGB point-source catalogue has been published from this
programme at the time of writing.
No suitable \textit{JWST} catalogue exists for our field at the time
of writing.

\textit{HST}/PHATTER \citep{Williams2021} covers our field in six
bands but provides no mid-infrared data.
\textit{WISE}/AllWISE \citep{Wright2010} has ${\sim}6\,\mathrm{arcsec}$
resolution, insufficient to resolve individual AGB stars at the
distance of~M33.

Of four Spitzer catalogues assessed, \citet{Thompson2009} showed a
systematic offset of $0.16$\,mag at $[3.6]$\,micron relative to the
Spitzer Enhanced Imaging Products (SEIP; the IRSA archival source
list combining IRAC and MIPS photometry across multiple programs),
attributable to a PSF-versus-aperture method difference, and was
not used.
\citet{Khan2015} produced unreliable cross-matches for
$\langle\mathrm{K}\rangle > 16\,\mathrm{mag}$ sources in the crowded
nuclear field (confirmed by low-amplitude variables yielding median
$\langle\mathrm{K}\rangle-[3.6] = 1.05\,\mathrm{mag}$,
physically implausible) and was excluded.
We adopt the SEIP as our primary Spitzer source, supplemented by
\citet{McQuinn2007}.

%----- Table 6 -------------------------------------------------------
\begin{table*}
\centering
\caption{Space-based photometric programmes evaluated for cross-matching
with the Paper~VII variable star catalogue
($r \leq 3\,\mathrm{arcmin}$ from the M33 nucleus,
$\alpha = 01^{\rm h}33^{\rm m}50\fs9$,
$\delta = +30\degr39\arcmin37\arcsec$, J2000).}
\label{tab:multiwave_summary}
\setlength{\tabcolsep}{4pt}
\begin{tabular}{lllccc}
\toprule
Programme & Reference & Instrument/Bands & Field & AGB cat. & Used \\
\midrule
\multicolumn{6}{l}{\textit{James Webb Space Telescope}} \\[1pt]
PHANGS-JWST (GO-2107) & \citet{Lee2023} & NIRCam+MIRI, $2$--$21\,\micron$
  & No (M33 excluded) & N/A & No \\
GO-2128 MIRI & \citet{Peltonen2024} & F560W, F2100W
  & No ($28\arcmin$ offset) & YSO & No \\
GO-2128 NIRCam & \citet{Sarbadhicary2025} & F090W, F200W, F335M, F444W
  & No (co-spatial) & SNR & No \\
GO-3436 & Rosolowsky (PI, 2024) & NIRCam F150W, F335M; MIRI F770W
  & Partial$^{d}$ & ISM/SNR & No \\[4pt]
\midrule
\multicolumn{6}{l}{\textit{Hubble Space Telescope}} \\[1pt]
PHATTER & \citet{Williams2021} & F275W--F160W
  & Full & Optical & $\langle\mathrm{J}\rangle$ only$^{a}$ \\[4pt]
\midrule
\multicolumn{6}{l}{\textit{WISE}} \\[1pt]
AllWISE & \citet{Wright2010} & W1--W4
  & Full & $6\arcsec$ res. & No \\[4pt]
\midrule
\multicolumn{6}{l}{\textit{Spitzer Space Telescope}} \\[1pt]
SEIP & \citet{SEIP2019} & IRAC $[3.6]$--$[8.0]\,\micron$
  & Full & Yes & Yes (primary) \\
McQuinn et al.\ (2007) & \citet{McQuinn2007} & IRAC $[3.6],[4.5],[8.0]\,\micron$
  & Full & Yes & Yes (secondary) \\
Thompson et al.\ (2009) & \citet{Thompson2009} & IRAC $[3.6],[4.5]\,\micron$
  & Full & Yes & No$^{b}$ \\
Khan et al.\ (2015) & \citet{Khan2015} & IRAC+MIPS
  & Full & Yes & No$^{c}$ \\
\bottomrule
\end{tabular}
\begin{minipage}{\linewidth}
\smallskip\small
$^{a}$~Mean $\langle\mathrm{J}\rangle$ photometry from \DIFdelbeginFL \DIFdelendFL \DIFaddbeginFL \DIFaddFL{Paper~I
}\DIFaddendFL used for the CMD (Fig.\,\ref{fig:cmd}); no mid-infrared data.\\
$^{b}$~Systematic offset $0.16\,\mathrm{mag}$ at $[3.6]\,\micron$ vs
SEIP (PSF vs aperture method); excluded to preserve homogeneity.\\
$^{c}$~Unreliable matches for $\langle\mathrm{K}\rangle>16\,\mathrm{mag}$
due to source confusion in the crowded nuclear field; validated and excluded.\\
$^{d}$~Central field covered but programme targets ISM and supernova
remnants; no AGB stellar point-source catalogue published at the
time of writing.
\end{minipage}
\end{table*}
%---------------------------------------------------------------------

\subsubsection{Cross-match with Spitzer IRAC}
\label{sec:spitzer_xmatch}

The SEIP Source List \citep{SEIP2019} was retrieved from IRSA via
cone search of radius $3\,\mathrm{arcmin}$, returning 1599~sources.
Fluxes were converted to Vega magnitudes using
$m = -2.5\log_{10}(F_\nu/F_0)$, with zero-points
$F_0([3.6]) = 280.9\,\mathrm{Jy}$ and $F_0([4.5]) = 179.7\,\mathrm{Jy}$
\citep{Reach2005}.
For stars without SEIP detections, we incorporated photometry from
\citet{McQuinn2007}; the two catalogues are mutually consistent to
$0.07 \pm 0.34\,\mathrm{mag}$ at $[3.6]$ on 94~common sources.
Seven sources have valid $[4.5]$ photometry but lack reliable $[3.6]$
flux in the SEIP pipeline (\texttt{i1\_fluxflag}~$= 2$), reflecting
band-dependent source confusion in the crowded nuclear field
\citep{Fazio2004}; these stars are included in the $[4.5]$ sample
but excluded from the $\langle\mathrm{K}\rangle-[3.6]$ dust
diagnostic.

All cross-matches used a reciprocal best-match algorithm with
$r < 1\,\mathrm{arcsec}$: a pair was accepted only if each source
was the other's nearest neighbour.
This radius is conservative relative to the SEIP positional accuracy
($\lesssim 0.3\,\mathrm{arcsec}$, median separation
$0.23\,\mathrm{arcsec}$) and well within the IRAC $[3.6]$ FWHM
of $1.66\,\mathrm{arcsec}$ \citep{Fazio2004}.
Extending the radius to $1.5\,\mathrm{arcsec}$ would add only
6~SEIP counterparts (separations $1.1$--$1.5\,\mathrm{arcsec}$)
while more than doubling the expected false-match rate
(from 42 to 94~sources); this was not adopted.
No genuine match falls between $0.85\,\mathrm{arcsec}$ and
$1.0\,\mathrm{arcsec}$, confirming the adopted radius does not
truncate any real counterparts.
Our $\langle\mathrm{K}\rangle$ magnitudes are on the MKO system
\citep{Tokunaga2002}; the colour transformation to the Spitzer Vega
system is $<0.05\,\mathrm{mag}$ for AGB stars \citep{Carpenter2001},
and the extinction correction
$A_K - A_{[3.6]} \approx 0.011\,\mathrm{mag}$ \citep{Cardelli1989}
is negligible; neither was applied.

%----- Table 7 -------------------------------------------------------
\begin{table}
\centering
\caption{Results of the Spitzer IRAC cross-match ($N=847$ stars).}
\label{tab:xmatch_results}
\begin{tabular}{lc}
\toprule
Quantity & Value \\
\midrule
SEIP matches (primary) & 114 \\
McQuinn et al.\ (2007) matches (secondary) & 42 \\
Total unique with $[3.6]$ & 156 \\
\quad With $[4.5]$ & 157 \\
\quad With $[8.0]$ & 62 \\
\quad With $[3.6]$ and $[4.5]$ & 150 \\
Stars with $\langle\mathrm{J}\rangle$ (\citealt{Javadi2011a}) & 826 \\
Overall match fraction & \DIFdelbeginFL \DIFdelendFL \DIFaddbeginFL \DIFaddFL{18}\DIFaddendFL ~per cent \\
\quad $\langle\mathrm{K}\rangle = 14$--16 & $\sim$60~per cent \\
\quad $\langle\mathrm{K}\rangle = 16$--17 & 28~per cent \\
\quad $\langle\mathrm{K}\rangle = 17$--18 & 9~per cent \\
Dust-enshrouded ($\langle\mathrm{K}\rangle-[3.6]>0.5$) & 120 (lower limit) \\
Median $\langle\mathrm{K}\rangle-[3.6]$ & 0.92~mag \\
\bottomrule
\end{tabular}
\end{table}
%---------------------------------------------------------------------

\subsubsection{Colour diagnostics and dust-enshrouded AGB candidates}
\label{sec:colour_diag}

\DIFdelbegin \DIFdelend \DIFaddbegin \DIFadd{We }\DIFaddend adopt
$\langle\mathrm{K}\rangle-[3.6] > 0.5\,\mathrm{mag}$ to classify
variable stars as dust-enshrouded AGB candidates.
Of the 156~stars with valid $[3.6]$ photometry, 120 (\DIFdelbegin \DIFdelend \DIFaddbegin \DIFadd{77}\DIFaddend ~per cent)
satisfy this criterion.
The $\langle\mathrm{K}\rangle-[3.6]$ index spans
$-0.35$ to $+3.44\,\mathrm{mag}$ with a median of
$0.92 \pm 0.70\,\mathrm{mag}$.
The $[3.6]-[4.5]$ index \DIFdelbegin \DIFdelend \DIFaddbegin \DIFadd{is }\DIFaddend available for 150~stars\DIFdelbegin \DIFdelend \DIFaddbegin \DIFadd{.
In the DUSTiNGS survey, extreme AGB (x-AGB) stars are those redder than
$[3.6]-[4.5] = 0.1\,\mathrm{mag}$ and brighter than
$M_{3.6} = -8\,\mathrm{mag}$ \citep{Boyer2015}; this colour does not by
itself determine whether the dust is carbon-rich or oxygen-rich}\DIFaddend .

\begin{figure*}
  \centering
  \includegraphics[width=\textwidth]{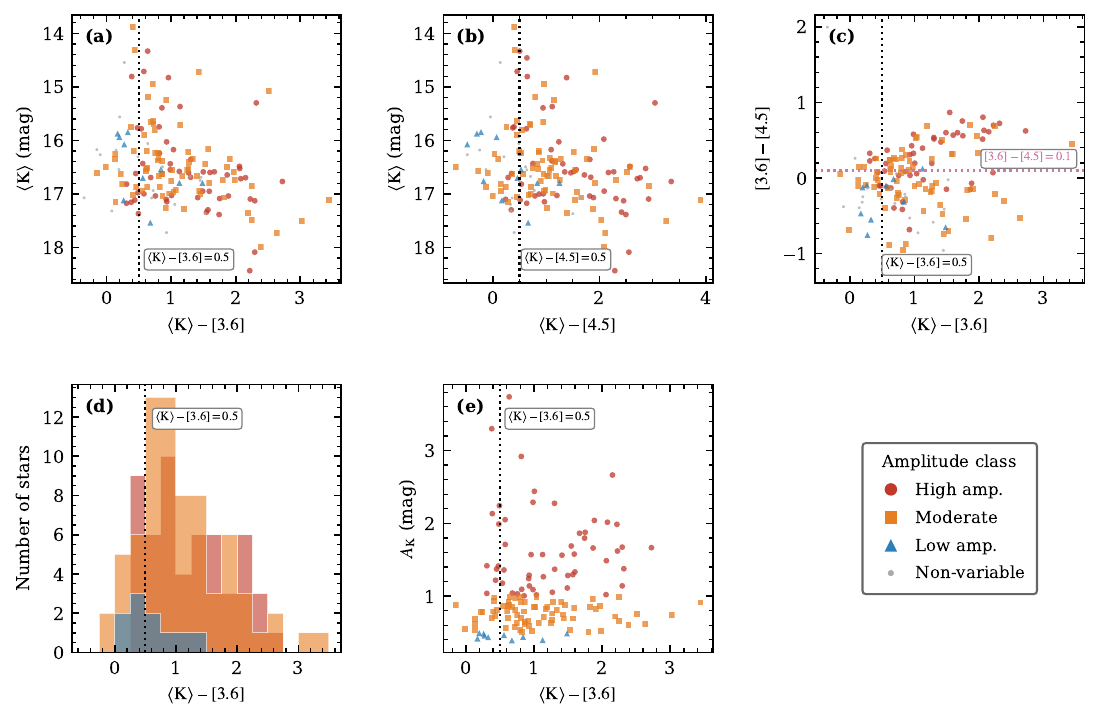}
  \caption{Multi-wavelength colour analysis for the Paper~VII
  variable star sample cross-matched with Spitzer IRAC photometry
  from the SEIP \citep{SEIP2019} (primary; 114~matches) and
  \citet{McQuinn2007} (secondary; 42~additional matches).
  (a)~Weighted-mean $\langle\mathrm{K}\rangle$ versus
  $\langle\mathrm{K}\rangle-[3.6]$ for 156~stars; the black dotted
  line marks the $\langle\mathrm{K}\rangle-[3.6] = 0.5\,\mathrm{mag}$
  dust-enshrouded threshold \DIFdelbeginFL \DIFdelendFL \DIFaddbeginFL \DIFaddFL{adopted in
  Section~\ref{sec:colour_diag}}\DIFaddendFL .
  (b)~$\langle\mathrm{K}\rangle$ versus
  $\langle\mathrm{K}\rangle-[4.5]$ for 157~stars; the black dotted
  line at $0.5\,\mathrm{mag}$ is shown for comparison with panel~(a).
  (c)~Colour--colour diagram $[3.6]-[4.5]$ versus
$\langle\mathrm{K}\rangle-[3.6]$ for 150~stars; the purple
dotted line at $[3.6]-[4.5] = 0.1\,\mathrm{mag}$ \DIFdelbeginFL \DIFdelendFL \DIFaddbeginFL \DIFaddFL{is the colour limit
used in the DUSTiNGS survey to define extreme AGB }\DIFaddendFL (\DIFdelbeginFL \DIFdelendFL \DIFaddbeginFL \DIFaddFL{x-AGB}\DIFaddendFL ) \DIFdelbeginFL \DIFdelendFL stars
\citep{Boyer2015}; the black dotted line \DIFdelbeginFL \DIFdelendFL \DIFaddbeginFL \DIFaddFL{is as in panel~(a)}\DIFaddendFL .
    (d)~Distribution of $\langle\mathrm{K}\rangle-[3.6]$ by
  pulsation amplitude class; the black dotted line \DIFdelbeginFL \DIFdelendFL \DIFaddbeginFL \DIFaddFL{is as in
  panel~(a)}\DIFaddendFL .
    (e)~Pulsation amplitude $A_K$ versus
  $\langle\mathrm{K}\rangle-[3.6]$ for the 136 confirmed variable
  stars with mid-infrared counterparts\DIFdelbeginFL \DIFdelendFL .
  In panels~(a)--(e), symbols are colour-coded by amplitude class:
  high ($A_{\mathrm{K}} \geq 1.0\,\mathrm{mag}$\DIFdelbeginFL \DIFdelendFL ), moderate
  ($0.5 \leq A_{\mathrm{K}} < 1.0\,\mathrm{mag}$\DIFdelbeginFL \DIFdelendFL ) \DIFdelbeginFL \DIFdelendFL and low
  ($A_{\mathrm{K}} < 0.5\,\mathrm{mag}$\DIFdelbeginFL \DIFdelendFL ).}
  \label{fig:multiwave}
\end{figure*}

The 120~candidates represent a \textit{strict lower limit}: of the
847~monitored stars, at least 120 (\DIFdelbegin \DIFdelend \DIFaddbegin \DIFadd{14}\DIFaddend ~per cent) are confirmed
dust-enshrouded AGB candidates; the \DIFdelbegin \DIFdelend \DIFaddbegin \DIFadd{18}\DIFaddend ~per cent Spitzer match
fraction is set by pipeline completeness in the crowded nuclear
field, not by the intrinsic dust fraction.
Even the brightest star in our sample
($\langle\mathrm{K}\rangle = 13.88\,\mathrm{mag}$) has
$V > 18.9\,\mathrm{mag}$ assuming $V-K = 5\,\mathrm{mag}$ for
O-rich AGB stars \citep{Javadi2011a}, already $1.9\,\mathrm{mag}$
below the ASAS-SN detection limit of $V < 17\,\mathrm{mag}$
\citep{Jayasinghe2018}.
Dust-enshrouded variables with $V-K \gtrsim 7$\,mag are
entirely beyond the reach of current optical surveys, confirming the
unique capability of our near-infrared monitoring programme
\citep{Javadi2011a, Javadi2015}.
Fig.~\ref{fig:multiwave} presents the resulting multi-wavelength
colour analysis.

\begin{figure}
  \centering
  \includegraphics[width=\columnwidth]{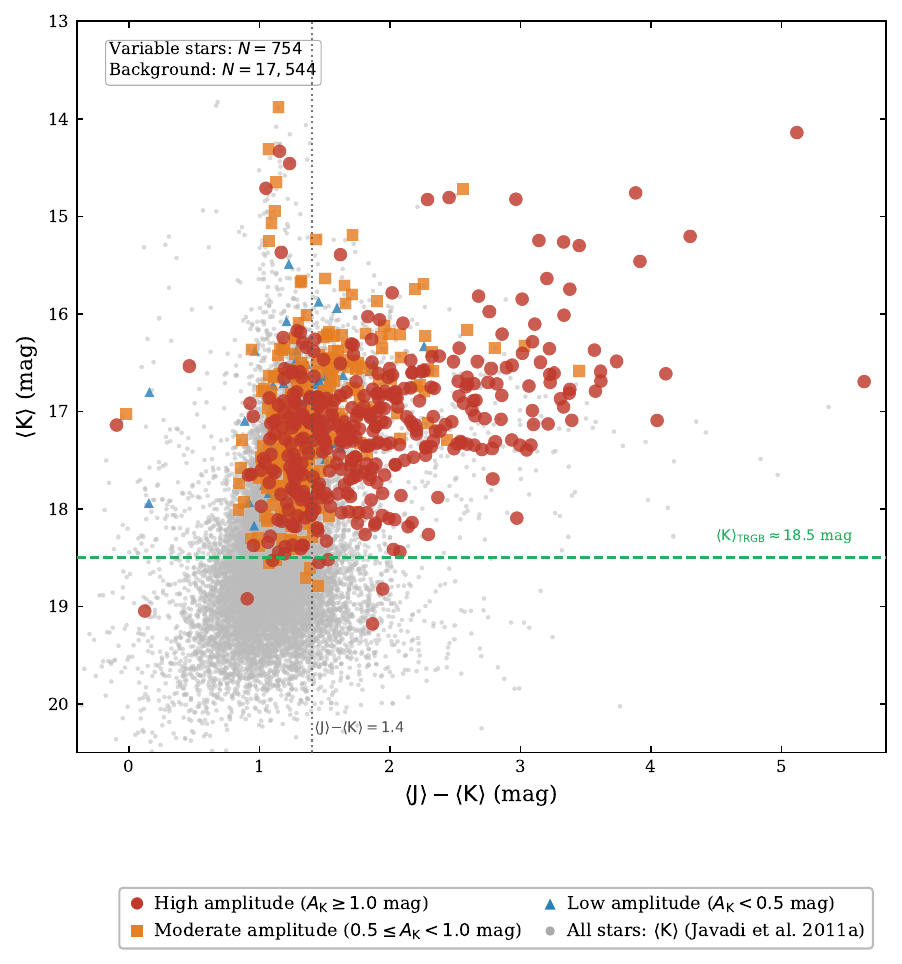}
  \caption{Near-infrared colour--magnitude diagram for stars in the
  central $3\,\mathrm{arcmin}$ of~M33\DIFdelbeginFL \DIFdelendFL \DIFaddbeginFL \DIFaddFL{: the }\DIFaddendFL background population
  \DIFdelbeginFL \DIFdelendFL \DIFaddbeginFL \DIFaddFL{from Paper}\DIFaddendFL ~\DIFaddbeginFL \DIFaddFL{I, and the }\DIFaddendFL Paper~VII catalogue stars with
  cross-matched $\langle\mathrm{J}\rangle$-band counterparts in
  \DIFdelbeginFL \DIFdelendFL \DIFaddbeginFL \DIFaddFL{Paper~I}\DIFaddendFL , colour-coded by pulsation amplitude class
  (see legend); the \DIFdelbeginFL \DIFdelendFL 72 \DIFdelbeginFL \DIFdelendFL non-variable \DIFdelbeginFL \DIFdelendFL stars \DIFaddbeginFL \DIFaddFL{among the latter }\DIFaddendFL are not
  distinguished from the \DIFdelbeginFL \DIFdelendFL background population.
  The horizontal axis shows
  $\langle\mathrm{J}\rangle - \langle\mathrm{K}\rangle$, where
  $\langle\mathrm{J}\rangle$ is the mean $J$-band magnitude from
  \DIFdelbeginFL \DIFdelendFL \DIFaddbeginFL \DIFaddFL{Paper~I }\DIFaddendFL (averaged over \DIFdelbeginFL \DIFdelendFL \DIFaddbeginFL \DIFaddFL{its }\DIFaddendFL 2003--2007 UKIRT
  monitoring baseline) and $\langle\mathrm{K}\rangle$ is our
  weighted-mean $K$ magnitude from the 2003--2018 baseline of
  Paper~VII; both datasets report mean magnitudes, but the
  differing 11-yr baselines introduce additional scatter in
  $\langle\mathrm{J}\rangle - \langle\mathrm{K}\rangle$ for
  large-amplitude pulsators.
  The horizontal dashed line marks \DIFdelbeginFL \DIFdelendFL \DIFaddbeginFL \DIFaddFL{the approximate TRGB level,
  $\langle\mathrm{K}\rangle \approx 18.5\,\mathrm{mag}$}\DIFaddendFL , and the
  dotted vertical line at
  $\langle\mathrm{J}\rangle - \langle\mathrm{K}\rangle =
  1.4\,\mathrm{mag}$ \DIFdelbeginFL \DIFdelendFL \DIFaddbeginFL \DIFaddFL{is }\DIFaddendFL the \DIFaddbeginFL \DIFaddFL{approximate colour divide between O-rich and
  C-rich }\DIFaddendFL AGB \DIFdelbeginFL \DIFdelendFL stars \DIFdelbeginFL \DIFdelendFL \DIFaddbeginFL \DIFaddFL{\citep[see][]{Cioni2006}.
  The }\DIFaddendFL reddest sources
  ($\langle\mathrm{J}\rangle - \langle\mathrm{K}\rangle \gtrsim
  3\,\mathrm{mag}$) correspond to the dust-enshrouded AGB candidates
  of Section~\ref{sec:colour_diag}.}
  \label{fig:cmd}
\end{figure}

Fig.\,\ref{fig:cmd} shows the near-infrared CMD for 826~matched
stars; \DIFdelbegin \DIFdelend \DIFaddbegin \DIFadd{742 of the 754 variables (98~per~cent) }\DIFaddend lie on the AGB sequence
above the \DIFdelbegin \DIFdelend \DIFaddbegin \DIFadd{approximate tip of the red giant branch (TRGB), consistent
with }\DIFaddend their evolved nature.
SED-based mass-loss rates will be presented in Paper~VIII.

%----- Figure 8 (6 panels) -------------------------------------------

%----- Figure 9 (CMD) ------------------------------------------------
%---------------------------------------------------------------------

%% ============================================================
%% SECTION 6 --- CONCLUSIONS
%% ============================================================

\section{Conclusions}
\label{sec:conclusions}

We have presented a homogenised K-band photometric catalogue for
the central kiloparsec of the Triangulum galaxy M33, combining
four PSF-based UKIRT datasets spanning 15.11~yr (2003--2018).
The principal results are as follows.

\textit{(i) Catalogue.}
Cross-matching four independent PSF-based near-infrared datasets
yields 847~unique stars with 21\,085~valid K-band measurements
spanning 15.11~yr, homogenised onto the 2MASS/Vega photometric
system.

\textit{(ii) PSF photometry improvements.}
We present, for the first time, \DIFdelbegin \DIFdelend PSF photometry for the
UFTI mosaic and new PSF reductions of WFCAM imaging for the four
central tiles, superseding earlier aperture photometry that
suffered from blending biases in the crowded nuclear field.

\textit{(iii) Temporal baseline.}
The combined 15.11-yr baseline is sufficient for reliable period
determination of Mira variables with $P \lesssim 2000$~d -- a
substantial improvement over the 3.80-yr baseline of the original
UIST survey alone.

\textit{(iv) Variable star census.}
Of 847 monitored stars, 771 (\DIFdelbegin \DIFdelend \DIFaddbegin \DIFadd{91}\DIFaddend ~per cent) are variable,
reflecting the luminosity-limited, AGB-dominated composition of
the sample. Among the variables, 385 (\DIFdelbegin \DIFdelend \DIFaddbegin \DIFadd{50}\DIFaddend ~per cent) have
peak-to-peak amplitudes $A_K \ge 1.0$~mag and are prime Mira-type
candidates.

\textit{(v) Multi-wavelength context.}
Cross-matching with archival mid-infrared photometry identifies
120 dust-enshrouded AGB candidates, a population that lies beyond
the reach of optical variability surveys.

\textit{(vi) Foundation for Paper~VIII.}
The Paper~VII catalogue provides the photometric and
multi-wavelength foundation for period determination,
period--luminosity analysis, dust mass-loss rate estimation, and
star formation history reconstruction, all to be presented in
Paper~VIII.

The full homogenised catalogue is available at the Centre de
Donn\'ees astronomiques de Strasbourg (CDS) upon acceptance.
%%
%% ============================================================
%% ACKNOWLEDGEMENTS
%% ============================================================
%%
\section*{Acknowledgements}
We thank the staff at UKIRT for their excellent support of this
programme. \DIFdelbegin \DIFdelend \DIFaddbegin \DIFadd{A.J.\ thanks the Leverhulme Trust }\DIFaddend for financial support
\DIFdelbegin \DIFdelend under grant No.\DIFaddbegin \DIFadd{\ }\DIFaddend RF/4/RFG/2007/0297, \DIFdelbegin \DIFdelend the Royal Astronomical Society,
and \DIFdelbegin \DIFdelend the Royal Society under grant No.\DIFaddbegin \DIFadd{\ }\DIFaddend IE130487.
UKIRT is owned by the University of Hawaii (UH) and operated by the
UH Institute for Astronomy. When the 2003--2007 data reported here
were obtained, UKIRT was operated by the Joint Astronomy Centre on
behalf of the Science and Technology Facilities Council of the
United Kingdom; when the 2018 UHS data were obtained, operations
were enabled through the cooperation of the East Asian Observatory.
This publication makes use of data from the Two Micron All Sky
Survey, a joint project of the University of Massachusetts and
IPAC/Caltech, funded by NASA and the NSF, and archival data from the
Spitzer Space Telescope and the NASA/IPAC Infrared Science Archive,
operated by JPL/Caltech under contract with NASA. \DIFdelbegin %

\DIFdelend \DIFaddbegin \DIFadd{Finally, we thank
the referee for her/his constructive report, which prompted us to
improve the manuscript.
}\DIFaddend %% ============================================================
%% DATA AVAILABILITY
%% ============================================================
%%
\section*{Data Availability}

The homogenised K-band \DIFdelbegin \DIFdelend \DIFaddbegin \DIFadd{summary catalogue (847 stars) and the
individual-epoch light curves (21\,085 measurements) }\DIFaddend described in
this paper will be made available at the Centre de Donn\'ees
astronomiques de Strasbourg (CDS) upon acceptance.
Reduced images and intermediate catalogues are available from the
corresponding author upon reasonable request.

\bibliographystyle{mnras}
\bibliography{paperVII}

\DIFaddend %% APPENDIX
%% ============================================================
%%

\FloatBarrier
\appendix

\section{Epoch count distribution}
\label{sec:appendix_epochs}

\noindent
\begin{minipage}{\columnwidth}
\captionof{table}{Distribution of K-band epoch counts across the
847-star master catalogue.}
\label{tab:epochs}
\centering
\begin{tabular}{lrr}
\hline
Epoch range & $N_\star$ & Per cent \\
\hline
4--10  &  17 &  2.0 \\
11--15 &  20 &  2.4 \\
16--20 & 346 & 40.9 \\
21--25 & 173 & 20.4 \\
26--30 &  89 & 10.5 \\
31--35 & 111 & 13.1 \\
36--40 &  41 &  4.8 \\
41--50 &  11 &  1.3 \\
51--60 &  33 &  3.9 \\
61--63 &   6 &  0.7 \\
\hline
Total  & 847 & 100  \\
\hline
\end{tabular}
\par\smallskip
\raggedright
\textit{Notes.} Percentiles: 10th $= 17$; 25th $= 19$; median $= 21$;
75th $= 30$; 90th $= 36$.
Maximum $= 63$ (Master~ID~518: 1 UFTI $+$ 49 UIST $+$ 13 WFCAM-PSF).
\end{minipage}

%% ============================================================
%% APPENDIX B — BORDERLINE VARIABLE STARS
%% ============================================================

\section{Borderline variable stars: single-epoch outlier analysis}
\label{sec:appendix_borderline}

Of the 771 stars classified as variable ($\sigma_{\mathrm{K}} > 0.10$~mag),
21 have $\sigma_{\mathrm{K}}$ values between 0.10 and 0.65~mag that are
dominated by a single deviant epoch: after iterative $3\sigma$-clipping,
their clipped scatter falls below the 0.10~mag threshold.
These stars are retained in the variable catalogue because (i)~their
peak-to-peak amplitude after outlier removal ($A_K^{\rm clip}$) remains
non-negligible (range 0.17--0.59~mag across all 21 stars;
Table~\ref{tab:borderline}), and (ii)~in 16 of the 21 cases the deviant
epoch has a photometric uncertainty that exceeds the star's median
uncertainty by a factor $\geq 1.5$, providing independent evidence of
a bad measurement rather than a genuine brightness excursion.

In 11 of the 21 cases (52~per cent) the outlier epoch originates from
the WFCAM-PSF dataset, consistent with the lower spatial resolution of
WFCAM ($0.40$~arcsec~pixel$^{-1}$) compared with UIST or UFTI and the
resulting higher susceptibility to blending in the crowded nuclear field.
The remaining cases are distributed among UHS (4), UFTI (4), and UIST (2).

The outlier epochs show a pronounced directional asymmetry: 18 of
21 (86~per cent) are fainter, not brighter, than the star's
clipped mean (median offset $-0.50$~mag). This is consistent with
instrumental or observing-condition artefacts (e.g.\ partial cloud,
poor seeing, residual blending), which preferentially reduce
measured flux, rather than genuine eclipses or flares, which would
show no directional preference.

All 21 stars are flagged in the published CDS catalogue with quality flag
\texttt{QF~=~1} (borderline variable; single-epoch outlier detected) to
allow users to apply stricter selection criteria if required.
Period analysis in Paper~VIII, using the full light curves with
epoch-quality weighting, will provide independent confirmation of
variability for these sources.

\begin{table*}
\centering
\caption{The 21 borderline variable stars whose $\sigma_{\mathrm{K}}$
exceeds the 0.10~mag variability threshold owing to a single deviant
epoch. $\sigma_{\mathrm{K}}^{\rm orig}$: standard deviation over all
epochs. $\sigma_{\mathrm{K}}^{\rm clip}$: standard deviation after
removing the $3\sigma$ outlier epoch. $A_K^{\rm clip}$: peak-to-peak
amplitude after outlier removal. $\sigma_{\rm out}/\sigma_{\rm med}$:
ratio of outlier epoch uncertainty to median epoch uncertainty; values
$>1.5$ indicate a likely bad epoch.}
\label{tab:borderline}
\setlength{\tabcolsep}{5pt}
\begin{tabular}{rccccccrl}
\hline
Master & $N_{\rm ep}$ &
$\sigma_{\mathrm{K}}^{\rm orig}$ &
$\sigma_{\mathrm{K}}^{\rm clip}$ &
$A_K^{\rm orig}$ &
$A_K^{\rm clip}$ &
$\sigma_{\rm out}/\sigma_{\rm med}$ &
Outlier & Verdict \\
ID & & (mag) & (mag) & (mag) & (mag) & & dataset & \\
\hline
166 & 17 & 0.169 & 0.091 & 0.729 & 0.352 &  3.2 & UHS       & Bad epoch (confirmed) \\
264 & 20 & 0.114 & 0.072 & 0.501 & 0.283 &  0.6 & UFTI      & Uncertain \\
286 & 20 & 0.114 & 0.079 & 0.530 & 0.281 &  4.9 & UHS       & Bad epoch (confirmed) \\
294 & 20 & 0.156 & 0.081 & 0.774 & 0.270 &  0.4 & UFTI      & Uncertain \\
349 & 28 & 0.122 & 0.097 & 0.577 & 0.436 &  1.7 & WFCAM-PSF & Bad epoch (likely) \\
390 & 27 & 0.282 & 0.066 & 1.542 & 0.310 &  1.8 & WFCAM-PSF & Bad epoch (likely) \\
443 & 32 & 0.241 & 0.088 & 1.104 & 0.343 &  1.8 & WFCAM-PSF & Bad epoch (likely) \\
448 & 26 & 0.114 & 0.082 & 0.562 & 0.282 &  0.9 & UFTI      & Uncertain \\
478 & 33 & 0.181 & 0.092 & 1.088 & 0.401 &  7.0 & WFCAM-PSF & Bad epoch (confirmed) \\
503 & 51 & 0.134 & 0.093 & 0.984 & 0.485 &  3.4 & UIST      & Bad epoch (confirmed) \\
520 & 32 & 0.647 & 0.041 & 3.738 & 0.172 &  2.8 & WFCAM-PSF & Bad epoch (likely) \\
540 & 31 & 0.134 & 0.090 & 0.789 & 0.594 &  4.7 & WFCAM-PSF & Bad epoch (confirmed) \\
556 & 29 & 0.116 & 0.080 & 0.628 & 0.397 &  1.4 & WFCAM-PSF & Uncertain \\
569 & 27 & 0.134 & 0.096 & 0.759 & 0.400 &  1.9 & UIST      & Bad epoch (likely) \\
597 & 18 & 0.132 & 0.089 & 0.581 & 0.295 &  1.7 & WFCAM-PSF & Bad epoch (likely) \\
666 & 15 & 0.265 & 0.087 & 1.205 & 0.360 &  1.6 & WFCAM-PSF & Bad epoch (likely) \\
689 & 14 & 0.403 & 0.067 & 1.598 & 0.236 & 12.2 & WFCAM-PSF & Bad epoch (confirmed) \\
693 & 28 & 0.165 & 0.098 & 0.869 & 0.383 &  8.9 & WFCAM-PSF & Bad epoch (confirmed) \\
696 & 17 & 0.138 & 0.078 & 0.643 & 0.249 &  4.8 & UHS       & Bad epoch (confirmed) \\
713 & 21 & 0.112 & 0.056 & 0.529 & 0.184 &  6.4 & UHS       & Bad epoch (confirmed) \\
722 & 19 & 0.216 & 0.083 & 1.073 & 0.318 &  0.6 & UFTI      & Uncertain \\
\hline
\end{tabular}
\begin{minipage}{\linewidth}
\smallskip
\textit{Notes.}
All 21 stars are flagged \texttt{QF~=~1} in the CDS catalogue.
$A_K^{\rm clip}$ uses the $3\sigma$-clipped light curve.
``Bad epoch (confirmed)'': $\sigma_{\rm out}/\sigma_{\rm med} > 3$;
the outlier epoch is almost certainly a photometric failure.
``Bad epoch (likely)'': $1.5 \leq \sigma_{\rm out}/\sigma_{\rm med} \leq 3$;
the elevated error suggests a compromised measurement.
``Uncertain'': $\sigma_{\rm out}/\sigma_{\rm med} < 1.5$; the epoch
appears photometrically normal; genuine low-level variability cannot
be excluded.
In all 21 cases, $A_K^{\rm clip} > 0.17$~mag, exceeding the
photometric noise floor; these stars are not classified as non-variable.
\end{minipage}
\end{table*}
\DIFdelbegin %

\DIFdelend

\label{lastpage}
\end{document}